\pdfoutput=1
\documentclass[nofootinbib]{revtex4}

\usepackage{graphicx}
\usepackage{color}
\usepackage{amsfonts}
\usepackage{amsmath}
\usepackage{slashed}
\def\la{\langle}
\def\ra{\rangle}

\def\l{\left}
\def\r{\right}

\def\beq{\begin{equation}}
\def\eeq{\end{equation}}
\def\bea{\begin{eqnarray}}
\def\eea{\end{eqnarray}}
\def\barr{\begin{array}}
\def\earr{\end{array}}

\begin{document}

\begin{titlepage}
\begin{center}
{\Large \bf Finite-size scaling for
four-dimensional Higgs-Yukawa model\\near the Gaussian fixed point} 
\vskip1cm {\large\bf
David Y.-J Chu$^{a}$,
Karl~Jansen$^{b}$, Bastian~Knippschild$^{c}$, C.-J. David~Lin$^{d,e}$}\\ \vspace{.5cm}
{\normalsize {\sl $a$ Department of Electrophysics, National Chiao-Tung
    University, Hsinchu 30010, Taiwan\\
$^{b}$ NIC, DESY Zeuthen, D-15738 Zeuthen, Germany\\
$^{c}$ HISKP, Universit\"{a}t Bonn, D-53115 Bonn, Germany\\
$^{d}$ Institute of Physics, National Chiao-Tung University, Hsinchu
30010, Taiwan\\
$^{e}$ Centre for High Energy Physics, Chung-Yuan Christian
University, Chung-Li 32023, Taiwan    
}}

\vskip1.0cm {\large\bf Abstract:\\[10pt]} \parbox[t]{\textwidth}{{
We analyse finite-size scaling behaviour of a four-dimensional Higgs-Yukawa model
near the Gaussian infrared fixed point.   Through improving the mean-field
scaling laws by solving one-loop renormalisation group equations,
the triviality property of this model can be manifested in
the volume-dependence of
moments of the scalar-field zero mode.
The scaling formulae for the moments are derived in this work with the
inclusion of the leading-logarithmic corrections.
To test these formulae, we confront them
with data from lattice simulations in a simpler model, namely the
O(4) pure scalar theory, and find numerical evidence of good agreement.  Our results of the
finite-size scaling can in principle be employed
to establish triviality of Higgs-Yukawa models, or to search for
alternative scenarios in studying their fixed-point structure, if sufficiently large lattices can be reached.
}}
\end{center}
\end{titlepage}


\section{Introduction}
\label{sec:intro}
With the progress of experiments at the Large Hadron Collider (LHC),
it is becoming essential to understand the validity of the Standard
Model (SM).  In particular, for the scalar sector of the SM
there is very strong evidence that it is 
trivial~\cite{Aizenman:1981zz, Frohlich:1982tw, Luscher:1987ay, Luscher:1987ek, Luscher:1988uq, Bernreuther:1987hv,Kenna:1992np,Gockeler:1992zj,Wolff:2009ke, Weisz:2010xx, Hogervorst:2011zw, Siefert:2014ela},
leading to the fact that the cut-off scale is indispensable, necessitating the emergence of new physics at a yet unknown energy scale.  Before
the discovery of the scalar resonance around 125 GeV, this triviality
property was used to predict an upper
bound for the Higgs-boson mass~\cite{Dashen:1983ts}
which then later could indeed be computed on the 
lattice \cite{Hasenfratz:1987eh,Kuti:1987nr,Luscher:1988gc}.  
Given the experimental measurement of the Higgs-boson mass, now it 
can be an ingredient for the investigation of the appearance of new
physics beyond the
SM~\cite{Branchina:2013jra,Gies:2013fua,Chu:2015nha,Brehmer:2015rna,Biekotter:2016ecg,Chu:2017vmc,Dawson:2017ksx}.

%
%

It has been a common strategy to extend the SM {\it via} introducing novel
interactions that involve relevant operators.  These relevant
operators can then result in non-trivial vacuum structure, leading to 
dynamical electroweak symmetry breaking (dEWSB).   For instance, in
the walking-technicolour scenario~\cite{Holdom:1984sk,Yamawaki:1985zg,Appelquist:1986an}, the SM Higgs sector is
replaced by a novel gauge theory coupled to technifermions.  
This new gauge coupling is asymptotically free.  It evolves very
slowly under renormalisation group (RG) transformation, and becomes strong at the
electroweak scale.  In such a scenario, the technifermion condensate
leads to dEWSB, and there can be a
light scalar bound state because of the quasi scale invariance. 
In the past decade, there have been many works to search for
candidate theories for such a scenario using lattice computations~\cite{Lucini:2015noa,Nogradi:2016qek,Svetitsky:2017xqk} and the gauge-gravity
duality~\cite{Piai:2010ma,Piai:2014pla}.  Another example is the class of theories
that are generically named the composite Nambu-Goldstone Higgs
models~\cite{Kaplan:1983sm,Georgi:1984af,Dugan:1984hq}, 
in which the Higgs particle is one of the Goldstone bosons in
a new strongly-coupled sector beyond the SM.  The Higgs boson
then acquires its
mass through the interaction between this new sector and the SM, which misaligns the vacuum and breaks the electroweak symmetry.   In
recent years, schemes for the ultraviolet (UV) completion of
such models have been proposed, see, {\it e.g.},
Refs.~\cite{Barnard:2013zea,Ferretti:2013kya,Cacciapaglia:2014uja,Belyaev:2016ftv}.
These proposals all involve asymptotically-free gauge theories, and
some of them have been studied numerically with lattice regularisation~\cite{DeGrand:2015lna,Ayyar:2017qdf,Ayyar:2018zuk,Ayyar:2018ppa,Bennett:2017kga,Lee:2018ztv}.

%
%

In the research activities described above, an important ingredient for finding physics 
beyond the SM is the knowledge of the fixed-point structure of candidate theories that can result in 
dEWSB.  Amidst the significant amount of projects searching for new, relevant
operators beyond the SM, we stress that our understanding for the non-perturbative aspect of the 
SM electroweak sector can still be improved.  While it is widely accepted that the 
O(4) scalar theory is trivial, where no relevant operator can emerge, the situation with the Higgs-Yukawa model
still requires further clarification.  
This is the main motivation of the current work, in which we develop a
strategy for studying the
fixed-point structure of the Higgs-Yukawa model by employing the 
technique of finite-size scaling (FSS). 
As pointed out in 
Refs.~\cite{Hung:2009hy}, perturbation theory
indicates that there can be interesting phase structures in the
Higgs-Yukawa model without gauge fields.  In particular, a
UV fixed point may be present.  In such a scenario,
the RG flows do not always
approach the Gaussian (mean-field) fixed point in the infrared (IR) regime, and the scalar quartic operator can
be relevant.  This UV fixed point typically appears at the values of the couplings where
perturbation theory may not be applicable~\cite{Molgaard:2014mqa}.
Therefore it would be necessary
to perform studies using nonperturbative approaches,
amongst which lattice field theory is the most reliable method.
Compared to the
pure scalar theories, lattice investigation of the Higgs-Yukawa model is much more challenging.  
In particular, it is conceptually difficult to regularise 
chiral fermions in Lattice Field Theory.  There were many
lattice calculations for the phase structure and the spectrum of the 
Higgs-Yukawa model without the continuum-like chiral symmetry around 1990.  See
Refs.~\cite{Shigemitsu:1987ei, Lee:1987eg, Lee:1988ut, Bock:1989ye, Hasenfratz:1989jr,
  Lee:1989xq, Lee:1989mi, Shigemitsu:1989xb, Abada:1990ds,
  Bock:1990tv, Bock:1991kn, Bock:1991bu, Bock:1991da,
  Hasenfratz:1991it, Kuti:1991vg, Hasenfratz:1992xs} for an incomplete
list of them, see Ref.~\cite{Maiani:1992pn} for a proceedings article
covering the attempts at that time. 
These early works led to evidence that the
phase structures of the Higgs-Yukawa model can be very rich.  
In the past decade, many aspects of
such studies have been extended to the case of chiral Yukawa 
couplings~\cite{Fodor:2007fn, Gerhold:2007yb, Gerhold:2007gx,
  Gerhold:2009ub, Gerhold:2010bh, Gerhold:2011mx, Bulava:2012rb}.   A
thorough, detailed analysis of the scaling behaviour of the theory
can provide a very useful tool to differentiate the scenarios sketched above.   We emphasise that only such an analysis can lead to
reliable results regarding the existence of relevant operators in the
Higgs-Yukawa model.

%
%

Finite-size technique is a well established and frequently used tool for studying the scaling
behaviour of a model~\cite{Fisher:1972zza}.   It also combines
naturally with lattice simulations.  In this approach, one first looks for
second-order phase transitions in the space of the
bare couplings of a theory.  Such phase transitions, when appearing at zero
temperature and infinite volume, are
the limit where the momentum cut-off approaches infinity.   That is, they
represent the continuum limit when the theory is regularised with the
space-time lattice.    In the vicinity of these transitions,  the
lattice size is the only available low-energy scale, and it can be regarded
as the renormalisation scale.    One can then use the RG equations
(RGE's) to investigate the correlators and obtain their dependence on
the finite size of the system.  In this procedure, the presence of either a Gaussian or a non-trivial fixed point is
crucial.  It ensures that the dimensionless couplings approach
constants, and the relevant couplings will combine with the lattice
volume to result in FSS behaviour.   As pointed out in 
Ref.~\cite{Brezin:1981gm}, care has to be taken when implementing this strategy in space-time dimension higher than three, because of 
issues arising from the properties of IR fixed points.
It was later demonstrated~\cite{Brezin:1985xx} that for correlators in four dimensions, 
logarithmic corrections to the volume-dependence play an essential role in the mean-field FSS behaviour.
This is actually a manifestation of the fact
that logarithmic scaling is a signature of Gaussian fixed points~\cite{Frohlich:1982tw}.
There have been studies on these
logarithms for four-dimensional scalar
models~\cite{Bernreuther:1987hv,Kenna:1992np,Gockeler:1992zj}.   These studies
resulted in evidence that such models are indeed trivial.  In the present work, we extend the FSS 
analysis to four-dimensional Higgs-Yukawa models, making thus a significant step beyond the existing 
results.  In fact, to the best of our knowledge, our work is the first to derive FSS formulae for the
Higgs-Yukawa models.

%
%

In this paper, we present results for our investigation of the corrections from leading logarithms (LL's) to the
mean-field scaling laws for moments of the scalar-field zero mode in a
four-dimensional Higgs-Yukawa model.
Our study has been performed  {\it
  \`{a}la } Brezin and Zinn-Justin~\cite{Brezin:1985xx}.   The main finding of this work is that the functional
form of these LL-improved mean-field scaling laws can be derived, and
it can be combined with lattice computations to probe the nature of
fixed points in the theory.  Following a scan of the bare parameter
space to search for critical points by employing numerical
simulations, one can then apply the strategy we develop here to
confirm triviality of the Higgs-Yukawa theory,
or to look for alternative scenarios.  As can be seen in Sec.~\ref{sec:scaling_formulae}, the logarithmic volume-dependence is quite complex.  
Therefore, to verify the results of our analytic calculation, we examine a simpler model, namely the O(4) pure scalar theory 
and confront the scaling formulae with numerical data from lattice simulations.  As shown in Sec.~\ref{sec:O4_numerical_test}, it is observed that our formulae fit the numerical data well.

%
%

This article is organised as the following.   In Sec.~\ref{sec:FSS_RG}
we briefly review the FSS technique and its
application near the mean-field fixed point, and discuss our
derivation of the scaling formulae for the Higgs-Yukawa model near the
Gaussian fixed point.  
Section~\ref{sec:O4_numerical_test} presents the numerical test by
confronting our scaling formulae in a simpler model, namely the O(4)
pure scalar theory, with lattice simulations.
Our conclusion is in Sec.~\ref{sec:conclusion}.

\section{Finite-size scaling and the renormalisation group}
\label{sec:FSS_RG}
The extraction of anomalous dimensions (critical exponents) is of
crucial importance for the study of universality classes in a field
theory.  For this purpose, the technique of FSS is the most commonly-employed
approach.   In the current project, our primary task is to establish FSS tools for
a class of Higgs-Yukawa models (described in Sec.~\ref{sec:HY_model}) near the Gaussian fixed point.
As explained in the rest of this section, this is achieved by extending the strategy developed for pure
scalar field theories in Ref.~\cite{Brezin:1985xx}.

\subsection{The Higgs-Yukawa model}
\label{sec:HY_model}
We work with the four-dimensional Euclidean Higgs-Yukawa
theory that contains one Dirac-fermion doublet,
\beq
\label{eq:fermion_doublet}
 \Psi = \left ( \begin{array}{c} t\\ b\end{array}\right ) \, ,
\eeq
and four real scalar fields, $\phi_{0,1,2,3}$, which can be
represented in the quaternion form,
\beq
\label{eq:quaternion_scalar}
 \Phi = \frac{1}{2}\sum_{\alpha=0}^{3} \theta_{\alpha} \phi_{\alpha}, \mbox{ }
 {\mathrm{where}}\mbox{ } \theta_{0} = {\mathbb{I}_{2\times 2}}, \mbox{}
 {\mathrm{and}} \mbox{ } \theta_{j} = i \sigma_{j}  \mbox{ } (j = 1,
 2, 3) \, ,
\eeq
with $\sigma_{j}$ being the Pauli matrices.
The Lagrangian of this model in Euclidean space in the continuum is 
\bea
\label{eq:HY_lagrangian}
 {\mathcal{L}}_{{\mathrm{HY}}} &=&  {\mathrm{Tr}} \left ( \partial_{\mu}
   \Phi^{\dagger}\partial_{\mu} \Phi \right )
+  M^{2}  {\mathrm{Tr}} \left ( \Phi^{\dagger} \Phi \right )
 + \lambda \left [  {\mathrm{Tr}} \left ( \Phi^{\dagger}
   \Phi \right ) \right ] ^{2} \nonumber \\
 && \hspace{3cm} +
  \bar{\Psi} \slashed\partial
    \Psi +  2  y   \left (
 \bar{\Psi}_{{\mathrm{L}}} \Phi \Psi_{{\mathrm{R}}}  + \bar{\Psi}_{{\mathrm{R}}} \Phi^{\dagger}
 \Psi_{{\mathrm{L}}}  \right ) \, ,
\eea
where $M$ is related to the Higgs-boson mass in the broken
phase of the model, $\lambda$ is the quartic self-coupling of the
scalar fields, and $y$ is the Yukawa coupling.  The pure scalar sector of the above model contains the usual
O(4) symmetry.   Notice that
we are setting the ``top'' and ``bottom'' Yukawa couplings to be
degenerate.   
Because of this degeneracy, the operators involving fermions are symmetric
under the transformation
\beq
\label{eq:LR_su2_symmetry}
 \Psi_{{\mathrm{L}}} \longrightarrow U_{{\mathrm{L}}} \Psi_{{\mathrm{L}}}, \mbox{ }\mbox{ } 
 \Psi_{{\mathrm{R}}} \longrightarrow U_{{\mathrm{R}}}\Psi_{{\mathrm{R}}} , \mbox{ }\mbox{ } 
 \Phi \longrightarrow U_{{\mathrm{L}}} \Phi U_{{\mathrm{R}}}^{\dagger}
 \, ,
\eeq
where $U_{{\mathrm{L}}}$ and $U_{{\mathrm{R}}}$ are elements of the left- and right-handed
SU(2) groups [${\mathrm{SU(2)}}_{{\mathrm{L}}}$ and ${\mathrm{SU(2)}}_{{\mathrm{R}}}$],
respectively.  Since 
${\mathrm{SU(2)}}_{{\mathrm{L}}}\times{\mathrm{SU(2)}}_{{\mathrm{R}}}$ covers the O(4)
group, the presence of the Yukawa interactions does not explicitly break the
O(4) symmetry in the scalar sector.   In addition to the above
symmetries, the Lagrangian in Eq.~(\ref{eq:HY_lagrangian}) is
invariant under
\beq
\label{eq:discrete_chiral_symmetry}
 \Psi \longrightarrow \gamma_{5} \Psi, \mbox{ }\mbox{ }  \Phi
 \longrightarrow -\Phi \, .
\eeq
This ``generalised-$Z_{2}$'' symmetry will play an important role in
deriving the scaling formulae, as explained in the next subsection.

%
%

Finally, for convenience, we also define 
\beq
\label{eq:Y_def}
 Y =  y^{2} \,  ,
\eeq
which exhibits the same classical scaling behaviour as
the quartic coupling, $\lambda$, in arbitrary space-time dimension.

\subsection{Derivation of finite-size scaling formulae near the
  Gaussian fixed point}
\label{sec:scaling_formulae}
To investigate universality classes of a theory, it is
crucial to understand the renormalisation group (RG) running behaviour
of its couplings in the vicinity of critical points.   In the study of
critical phenomena, such scaling
behaviour reveals the character of the associated fixed point.
For four-dimensional quantum field theories, a special feature of the Gaussian fixed point
is the presence of a double zero in the beta function.  This feature
results in logarithmic scaling behaviour \cite{Frohlich:1982tw}.
In this subsection, we will demonstrate that one can derive mean-field
FSS formulae for Higgs-Yukawa model, and
the leading logarithms can be obtained by employing one-loop
perturbation theory.   These logarithms then result in corrections to the
mea-field scaling laws.

%
%

In our calculations, we analyse the theory in
Eq.~(\ref{eq:HY_lagrangian}) on an anisotropic box with the four-volume being $L^3 \times L_t$, 
where $L_t$ is the Euclidean temporal extent.   We define the
anisotropy in lattice size,
\beq
\label{eq:s_def}
 s \equiv \frac{L_{t}}{L} \, ,
\eeq
which is kept constant in our calculation.

We first briefly review the generic argument for FSS,
proceeding with the theory described by the Lagrangian in Eq.~(\ref{eq:HY_lagrangian}).
For this purpose, we investigate a bare matrix element, 
${\mathcal{M}}_b [ M_{b}^{2} , \lambda_{b},  Y_{b}; a, L ]$
of classical (mass) dimension $D_{{\mathcal{M}}}$, where all the external momenta are
vanishing.  This matrix element depends on the
couplings, $M_{b}^{2}$, $\lambda_{b}$ and $Y_{b}$, which are the bare counterparts
of the quadratic, quartic and the Yukawa couplings in
Eqs.~(\ref{eq:HY_lagrangian}) and (\ref{eq:Y_def}).  We envisage that
${\mathcal{M}}_{b}$ is computed using lattice regularisation.  Therefore it is also a
function of the lattice spacing, $a$, and size, $L$.   One obtains
the corresponding renormalised matrix
element in a scheme {\it via}
\beq
\label{eq:ME_renorm}
 {\mathcal{M}} [ M^{2}(l) , \lambda(l),  Y(l); l, L ]  =
 Z_{{\mathcal{M}}} \left (
 l/a \right ) \times  {\mathcal{M}}_b [ M_{b}^{2} , \lambda_{b}, 
Y_{b}; a, L ] \, ,
\eeq
where $l$ is the renormalisation (length-)scale, and the matching
coefficient, $Z_{{\mathcal{M}}}$, can usually be determined either numerically or
analytically.  For convenience, in the discussion below we rescale
all dimensionfull
quantities by appropriate powers of a common scale, and denote their
corresponding, rescaled, dimensionless counterparts with a caret on
top.     Natural
candidates for this common scale can be the lattice spacing, $a$, and
the renormalisation scale, $l$, in Eq.~(\ref{eq:ME_renorm}).   

%
%

To observe the FSS behaviour of
$\hat{{\mathcal{M}}}_b$, one first performs the RG running from the
renormalisation scale $l$ to $L$ for $\hat{{\mathcal{M}}}$ in Eq.~(\ref{eq:ME_renorm}).  This leads to
\bea
\label{eq:RG_and_FSS_step1}
Z_{{\mathcal{M}}} (
 \hat{l}/\hat{a} ) \times  \hat{ {\mathcal{M}}}_b [ \hat{M}_{b}^{2} , \lambda_{b},  
Y_{b}; \hat{a}, \hat{L} ] &=&  \hat{{\mathcal{M}}} [ \hat{M}^{2}(l) , \lambda(l), 
Y(l); \hat{l}, \hat{L} ]  \nonumber\\
&& \nonumber\\
 &=& \zeta_{{\mathcal{M}}} (\hat{l},\hat{L}) \hat{L}^{-D_{{\mathcal{M}}}}  \hat{{\mathcal{M}}} [ \hat{M}^{2}(L) \hat{L}^{2} , \lambda(L), 
Y(L); 1, 1 ] \, , 
\eea
where
\beq
\label{eq:zeta_M}
 \zeta_{{\mathcal{M}}} (\hat{l},\hat{L}) = \exp \left( \int_{\hat{l}}^{\hat{L}}
   \gamma_{{\mathcal{M}}} (\rho) \mbox{ }d \log \rho \right) \, ,
\eeq
with $\gamma_{{\mathcal{M}}}$ being the anomalous dimension of 
${\mathcal{M}}$.   Assume that there exists a
strongly-coupled fixed point in the theory.  When the model is near
the critical surface of this fixed point at the length scale $L$, the renormalised dimensionless couplings, $\lambda(L)$
and $Y(L)$, as well as $\gamma_{{\mathcal{M}}}$ approach
constants.  This results in the scaling formula
\beq
\label{eq:FSS_strong_coupling}
 Z_{{\mathcal{M}}} (
 \hat{l}/\hat{a} ) \times  \hat{ {\mathcal{M}}}_b [ \hat{M}_{b}^{2} , \lambda_{b}, 
Y_{b}; \hat{a}, \hat{L} ] \times \hat{L}^{D_{{\mathcal{M}}} -
\gamma_{{\mathcal{M}}}} = f_{{\mathcal{M}}} [ \hat{M}^{2}(L)
\hat{L}^{2} ] \, , 
\eeq
where
\beq
\label{eq:M_L_and_M_l}
\hat{M}^{2}(L) = \hat{M}^{2}(l)
 \left ( \frac{L}{l} \right )^{-2 + 1/\nu_{M}} \, ,
\eeq
with $\nu_{M}$ being the anomalous dimension of this quadratic
coupling.  The above equation states that in the critical regime, the function, $f_{{\mathcal{M}}}$, depends on only one
parameter for all values of $\hat{L}$.  This is the usual FSS
behaviour.  Generically, it is not possible to derive the
explicit functional form of $f_{{\mathcal{M}}}$ for strongly-coupled
fixed points.  Nevertheless, in this case numerical studies for the
scaling properties in Eq.~(\ref{eq:FSS_strong_coupling}) can proceed with the use of
the curve-collapse method~\cite{0305-4470-34-33-302}.  


%
%

It was pointed out by Brezin~\cite{Brezin:1981gm} that 
care has to be taken when applying 
the above
simple argument for critical points associated with Gaussian fixed
points.  As discussed in
Ref.~\cite{Brezin:1985xx},  the incorporation of logarithmic corrections
is crucial for deriving mean-field FSS laws in four space-time
dimensions.  The authors of Ref.~\cite{Brezin:1985xx} developed the
detailed techniques for investigating scaling behaviour for pure scalar
field theories.  In the current work, we extend their results for the
first time to the case
of the Higgs-Yukawa model of Eq.~(\ref{eq:HY_lagrangian}).   Below we
also demonstrate that near the Gaussian fixed point, explicit scaling
functions [$f_{{\mathcal{M}}}$ in Eq.~(\ref{eq:FSS_strong_coupling})] 
can be obtained at the accuracy of leading logarithms.

Consider the Euclidean partition function of the model in Eq.~(\ref{eq:HY_lagrangian})
\begin{equation}
\label{full_partition_function}
 Z_{{\mathrm{HY}}}=\int {\mathcal{D}}\Phi \mbox{ }
 {\mathcal{D}}\bar{\Psi} \mbox{ } {\mathcal{D}} \Psi \exp
 \left \{  -S_{{\mathrm{HY}}}[\Phi, \bar{\Psi}, \Psi] \right \} \, , 
\eeq
where
\beq
 S_{{\mathrm{HY}}} = \int d^{4}x \mbox{ }{\mathcal{L}}_{{\mathrm{HY}}}
 \, ,
\end{equation}
with $\Psi$ and $\Phi$ defined in Eqs.~(\ref{eq:fermion_doublet}) and
(\ref{eq:quaternion_scalar}).  To proceed, we first integrate out the
fermion fields.  This results in
\beq
\label{eq:partition_function_int_out_fermions}
 Z_{{\mathrm{HY}}}=\int {\mathcal{D}}\Phi \exp \left \{  -
     \tilde{S}_{{\mathrm{eff}}} [\Phi
 ] \right \} \, ,
\eeq
and $\tilde{S}_{{\mathrm{eff}}}$ includes the effects from the
fermion determinant.  It is straightforward, with an explicit
calculation, to show that 
\beq
\label{eq:tilde_S_eff_Phi_square_form}
 \tilde{S}_{{\mathrm{eff}}} \left [ \Phi \right ] 
 = Z_{\Phi}
 {\mathrm{Tr}} \left ( \partial_{\mu} \Phi^{\dagger} \partial_{\mu}
   \Phi \right ) + 
 \tilde{V}_{{\mathrm{eff}}} \left [ \Phi \right ] \, , 
\eeq
where
\beq
\label{eq:V_eff_full}
\tilde{V}_{{\mathrm{eff}}} \left [ \Phi
 \right ] = \sum_{n=1}^{\infty} g_{2n}   \left [ {\mathrm{Tr}} \left (
 \Phi^{\dagger} \Phi \right ) \right ]^{n} \, ,
\eeq
with $Z_{\Phi}$ and  $g_{n}$ being functions of $M^{2}$, $\lambda$ and $Y$.   The
fact that $\tilde{V}_{{\mathrm{eff}}}$ contains only polynomials of $\left [ {\mathrm{Tr}} \left (
 \Phi^{\dagger} \Phi \right ) \right ]$ is a consequence of the
symmetries described in Eqs.~(\ref{eq:LR_su2_symmetry}) and
(\ref{eq:discrete_chiral_symmetry}).   For the current work, we are interested
in investigating this theory near the critical surface of the Gaussian fixed
point, where the scaling behaviour only receives logarithmic
corrections.  In this case, the operators with dimension larger than
four can be neglected in $\tilde{V}_{{\mathrm{eff}}}$.    That is,
this potential can be well approximated by
\beq
\label{eq:tilde_V_approx}
 \tilde{V}_{{\mathrm{eff}}} \left [ \Phi \right ] \approx g_{2}
 {\mathrm{Tr}} \left ( \Phi^{\dagger} \Phi \right ) + g_{4}
 \left [  {\mathrm{Tr}} \left ( \Phi^{\dagger} \Phi \right ) \right
 ]^{2} \, .
\eeq
%

%
%

Our goal is to study FSS behaviour of correlators involving only
the zero modes of the scalar fields in this theory.  We denote these
modes by $\varphi_{\alpha}$ ($\alpha = 0, 1, 2, 3$).  They are defined through
\beq
\label{eq:varphi_def}
 \varphi_{\alpha} = \frac{1}{V} \int_{V} d^{4} x \mbox{ }
 \phi_{\alpha} \, .
\eeq
Furthermore, we can perform the decomposition
\beq
\label{eq:phi_decomposition}
 \phi_{\alpha} = \varphi_{\alpha} + \chi_{\alpha} \, , 
\eeq
with
\beq
\int d^{4}
 x \mbox{ }\chi_{a} = 0 \, .
\eeq
That is, $\chi_{\alpha}$ are the fluctuations around the zero modes.  Near
the Gaussian fixed point, these fluctuations can be treated
perturbatively.  In this project, we work with one-loop precision in
deriving the FSS formulae.   It is straightforward to show that to
this order, the contributions to the partition function, $Z_{{\mathrm{HY}}}$, from terms that are quadratic and
non-quadratic in $\chi_{a}$ are completely factorised.   Together with
Eqs.~(\ref{eq:partition_function_int_out_fermions}),
(\ref{eq:tilde_S_eff_Phi_square_form}) and (\ref{eq:tilde_V_approx}),
this factorisation allows us to obtain the partition function at
one-loop,
\beq
\label{partition_function_after_zero_mode_separation}
 Z_{{\mathrm{HY}}}^{{\mathrm{1-loop}}}=\int_{-\infty}^{\infty} d^4 \varphi_{\alpha} \, {\mathcal{R}}_{\chi} \exp (-S_{{\mathrm{eff}}}[\varphi_{\alpha}])
 = {\mathcal{R}}_{\chi}  \Omega_{3} \int_{0}^{\infty} d \varphi \,
 \varphi^{3}  \exp (-S_{{\mathrm{eff}}}[\varphi]) \, , 
\eeq
where ${\mathcal{R}}_{\chi}$ is the non-quadratic contribution of $\chi_a$, 
$S_{{\mathrm{eff}}}$ is the effective action resulting from  
integrating out fermion fields and the bosonic degrees of freedom that are
quadratic in $\chi_a$, the symbol $\Omega_{3}$ stands for the
four-dimensional solid angle, and $\varphi$ is defined as
\beq
\label{eq:mod_varphi_def}
\varphi \equiv \sqrt{\sum_{\alpha=0}^{3} \varphi_{\alpha}^{2}} \, .
\eeq
%
In deriving Eq.~(\ref{partition_function_after_zero_mode_separation}),  we 
use the factorisation of ${\mathcal{R}}_{\chi}$ and the quadratic
terms in $\chi_{a}$ at one-loop order.  This allows us to treat ${\mathcal{R}}_{\chi}$ as an overal normalisation
constant, since we are only interested in obtaining scaling formulae
for moments of $\varphi$ at
the one-loop precision level.  In the rest of this
article, ${\mathcal{R}}_{\chi}$ is set to be unity.  On the other hand, the path integral over the fermionic
degrees of freedom, and the quadratic contributions from $\chi_{a}$
result in effects of renormalising the coupling constants in
$S_{{\mathrm{eff}}}$.  Near the critical surface of the Gaussian fixed
point, effects of this one-loop renormalisation can be studied
perturbatively with a saddle-point expansion around the zero
mode~\cite{Brezin:1985xx}.    It is straightforward to demonstrate that
this leads to
\begin{equation}
\label{effective_action_first_expansion_to_mean_field}
 \exp ( -S_{{\mathrm{eff}}}[\varphi] ) =  \exp \left( -s L^4M^{2}(r)
   \varphi^2 -s L^4\lambda (r) \varphi^4 \right) \, , 
\end{equation}
where $M^{2}(r)$ and $\lambda (r)$ are the one-loop renormalised
couplings, with $r$ being the renormalisation (length-)scale.

%
%

To proceed with the discussion of the FSS behaviour as governed by the
Gaussian fixed point, we first perform the change of variable,
\begin{equation}
\label{change_of_variable}
 \varphi = \left [ s L^4\lambda(r) \right ]^{-1/4}
 \check{\varphi} \, .
\end{equation}
This enables us to express the partition function in terms of
renormalised quantities as~\cite{Brezin:1985xx},
\beq
\label{partition_function_z_r}
 Z_{{\mathrm{HY}}}
^{{\mathrm{1-loop}}}= \Omega_{3} \left [  s
  L^4\lambda(r)\right ]^{-1} \int_0^{\infty} d\check{\varphi} \, \check{\varphi}^{3} 
 \exp \left(-\frac{1}{2}z_{r}\check{\varphi}^2 - \check{\varphi}^4
  \right) \, , 
\eeq
where
\beq
\label{eq:z_r_def}
z_{r} = \sqrt{s}\hat{L}^2\hat{M}^2(r) \lambda(r)^{-1/2} \, .
\eeq
Notice that the dimensionless zero-mode variable, $\check{\varphi}$, is obtained by rescaling $\varphi$ with
the lattice size.  In the limit where $L$ is large compared with all
the other scales in the theory, this also justifies the approximation
of the effective potential, $\tilde{V}_{{\mathrm{eff}}}$, in
Eq.~(\ref{eq:tilde_V_approx}).  

%
%

In order to derive the FSS behaviour of the
theory near the Gaussian fixed point, we resort to the general
renormalisation-group consideration summarised in
Eqs.~(\ref{eq:RG_and_FSS_step1}) and (\ref{eq:FSS_strong_coupling}).
Our aim is to investigate the scaling laws for moments of the
zero-mode variable, $\varphi$, which is also renormalisation-scale dependent. 
The first step is to identify the above renormalisation scale,
$r$, as the lattice size, $L$.   This results in 
\beq
\label{partition_function_scaling_variable}
 Z_{{\mathrm{HY}}}^{{\mathrm{1-loop}}} = \Omega_{3}  \left [ s L^4\lambda(L) \right ]^{-1}
 \bar{\varphi}_{3} (z), \mbox{ }{\mathrm{and}}\mbox{ }
 \la \varphi^{k} (L) \ra = \left [ s L^4\lambda(L) \right ]^{-k/4}
 \left [
 \frac{\bar{\varphi}_{k+3} (z)}{\bar{\varphi}_{3} (z)} \right ] \, ,
\eeq
with
\beq
\label{eq:scaling_var_def}
z \equiv z_{\hat{r}}|_{r=L} = \sqrt{s}\hat{L}^2\hat{M}^2(L)
\lambda(L)^{-1/2} \, ,
\eeq
and 
\beq
\label{eq:varphibarn_def}
  \bar{\varphi}_{n}(z) \equiv \int_{0}^{\infty} d\check{\varphi} \,
  \check{\varphi}^{n}  \exp \left(-\frac{1}{2}z\check{\varphi}^2 -
    \check{\varphi}^4 \right) \, .
\eeq
Equations~(\ref{partition_function_scaling_variable}) and (\ref{eq:scaling_var_def}) also
demonstrate the failure of the naive FSS argument for the Gaussian
fixed point, at which the quartic coupling vanishes.  One important step for practical implementation of the strategy
in Eqs.~(\ref{eq:RG_and_FSS_step1}) and (\ref{eq:FSS_strong_coupling})
is the matching of the bare lattice quantities to a renormalisation
scheme at a length scale $l$, followed by the renormalisation-group
running to $L$.    For the $k{-}$th moment of $\varphi$ in
Eq.~(\ref{partition_function_scaling_variable}), this means ($\delta_{{\mathcal{\varphi}}}$ is the anomalous dimension for
$\varphi$),
\bea
\label{eq:varphi_RG_running}
 \bar{v}_{k} \equiv \left [ s L^4\lambda(L) \right ]^{k/4} \la \varphi^{k}
(L) \ra &=&\left [ s \hat{L}^4\lambda(L) \right ]^{k/4} \la
\hat{\varphi}^{k}
(L) \ra
\nonumber\\
 &=&\left [ s \hat{L}^4\lambda(L) \right ]^{k/4} \exp \left( k \int_{\hat{l}}^{\hat{L}}
   \delta_{{\mathcal{\varphi}}} (\rho) \mbox{ }d \log \rho \right) \la \hat{\varphi}^{k}
(l) \ra  
\eea
is a function with only one argument, $z$.  That is, once a matching
scale, $l$ is chosen, and the running is performed, then $\bar{v}_{k}$
extracted at different lattice sizes will fall on a universal curve when plotted against $z$.

%
%

Before presenting further discussion for the details of the renormalisation-group running and
the logarithmic volume corrections in $z$ and $\varphi$, we find that the integrals in Eq.~(\ref{eq:varphibarn_def}) can be performed analytically, and the
results are
\begin{eqnarray}
\label{recursion_formula}
 \bar{\varphi}_0 &=& \frac{\pi}{8} \exp \left( \frac{z^2}{32} \right) \sqrt{|z|} 
 \left[ I_{-1/4}\left( \frac{z^2}{32} \right) - \mathrm{Sgn}(z) \,
                     I_{1/4}\left( \frac{z^2}{32} \right)  \right] \, ,\nonumber \\
 \bar{\varphi}_1 &=& \frac{\sqrt{\pi}}{8} \exp \left( \frac{z^2}{16} \right) 
 \left[ 1-\mathrm{Sgn}(z) \, \mathrm{Erf} \left( \frac{|z|}{4} \right)
                     \right] \, , \mbox{ } 
  \nonumber \\
 \bar{\varphi}_{n+2} &=& -2 \frac{d}{d z} \bar{\varphi}_n \, , 
\end{eqnarray}
where $I_{\nu}$ is the modified Bessel function of the first kind.
In other words, we have discovered that, unlike the generic case for
strongly-coupled fixed points, the FSS laws for the Gaussian fixed
point can be derived explicitly\footnote{One essential ingredient for
  making this statement is the truncation of
  $\tilde{V}_{{\mathrm{eff}}}$ in Eq.~(\ref{eq:tilde_V_approx}).  This
truncation may not be applicable for strongly-coupled fixed points,
because of the possibility for generating large anomalous dimensions
which can qualitatively alter the scaling behaviour of operators.}.  
To our knowledge, this work is the first derivation of
  the scaling behaviour encoded in 
  Eqs.~(\ref{eq:varphi_RG_running}) and (\ref{recursion_formula}), 
  although similar formulae for describing the O(4) pure-scalar theory in the large-volume
  limit were presented in Ref.~\cite{Gockeler:1992zj}.

%
%

With the results in Eq.~(\ref{recursion_formula}), in principle one
can obtain FSS behaviour for any moment of $\varphi$.   To use these
formulae for investigating the fixed-point structure of the theory
through lattice calculations, it is necessary to perform RG running between the
matching scale and the lattice size [Eqs.~(\ref{eq:RG_and_FSS_step1})
and (\ref{eq:FSS_strong_coupling})].   This 
can be achieved using one-loop perturbation theory, 
given that the FSS study is carried out in the vicinity of the
critical surface associated with the Gaussian fixed point\footnote{Notice that
  Eq.~(\ref{partition_function_after_zero_mode_separation}) is valid
  also only at one-loop.}.  The use of perturbation theory introduces
logarithmic corrections, as well as unknown parameters, to these scaling formulae.
To see explicitly how this is implemented, let us begin with the structure of
the one-loop renormalisation group equations for the theory in Eq.~(\ref{eq:HY_lagrangian}),
\begin{eqnarray}
 -\rho \frac{d}{d\rho}Y(\rho) &=& \beta_{YY^2} Y(\rho)^2 \, , 
  \nonumber \\
 -\rho \frac{d}{d\rho}\lambda(\rho) &=& \beta_{\lambda \lambda^2}\lambda(\rho)^2 + 
 \beta_{\lambda \lambda Y}\lambda(\rho)Y(\rho) + \beta_{\lambda Y^2}
                                        Y(\rho)^2 \, ,  \nonumber \\
  -\rho \frac{d}{d\rho}M^2(\rho) &=& 2\left[ \gamma_{Y}Y(\rho) +
    \gamma_{\lambda} \lambda(\rho) \right]M^2(\rho) \, , \nonumber\\
\label{mass_square_RGE}
-\rho \frac{d}{d\rho}\varphi(\rho) &=& 2\delta_{Y}Y(\rho)\varphi(\rho)
                                       \, ,
\end{eqnarray}
where the anomalous dimensions, $\beta$'s, $\gamma$'s and $\delta_{Y}$ are independent of
renormalisation scheme at this order.  We find that their values are 
\bea
\label{eq:one_loop_coeff_values}
 && \beta_{YY^{2}} = \frac{1}{\pi^{2}},\mbox{
 }\beta_{\lambda\lambda^{2}} = \frac{6}{\pi^{2}}, 
\mbox{ }\beta_{\lambda\lambda Y} = \frac{1}{\pi^{2}},\mbox{
}\beta_{\lambda Y^{2}} = -\frac{1}{4\pi^{2}} \, ,\nonumber\\
 && \gamma_{\lambda} = \frac{3}{2\pi^{2}},\mbox{ }\gamma_{Y} =
 \frac{1}{4\pi^{2}}, \mbox{ }
 \delta_{Y} = -\frac{1}{8\pi^{2}} \, .
\eea
To complete the discussion of the strategy for deriving FSS formulae,
we integrate the above RG equations from the scale $l$ to
$L$.  This results in the scaling variable, 
\begin{eqnarray}
\label{full_scaling_variable}
 z &=& \left( \frac{4\beta_{\lambda
       \lambda^2}}{    Y(l)       } \right)^{1/2} 
 \left[ Y(l)(\beta_{+}-\beta_{-}) 
 \right]^{\frac{2\gamma_{\lambda}}{\beta_{\lambda \lambda^2}}} \hat{L}^2
 \hat{M}^2(l) \nonumber\\ && \hspace{2 cm}
 \times \left[ 1 + Y(l) \beta_{YY^{2}} {\mathrm{log}} \l
   (   \frac{\hat{L}}{\hat{l}}\r )
 \right]^{\frac{1}{2}-\frac{2\gamma_Y}{\beta_{YY^2}}-\frac{\beta_{-}\gamma_{\lambda}}{\beta_{YY^2}\beta_{\lambda\lambda^2}}} 
\nonumber\\ && \hspace{2 cm}
 \times \frac{ \left\{ B_{+} -B_{-}\left[ 1 + Y(l) \beta_{YY^{2}} {\mathrm{log}} \l
   (   \frac{\hat{L}}{\hat{l}}\r )\right]^{\frac{\beta_{+}-\beta_{-}}{2\beta_{YY^2}}}  \right\}
 ^{\frac{1}{2}-\frac{2\gamma_{\lambda}}{\beta_{\lambda \lambda^2}}} 
 }{
 \left\{ \beta_{-}B_{+} -\beta_{+}B_{-}\left[ 1 + Y(l) \beta_{YY^{2}} {\mathrm{log}} \l
   (   \frac{\hat{L}}{\hat{l}}\r ) \right]^{\frac{\beta_{+}-\beta_{-}}{2\beta_{YY^2}}} 
 \right\}^{\frac{1}{2}}} \, ,
\end{eqnarray}
where 
\bea
\beta_{\pm} &=& (\beta_{YY^2}-\beta_{\lambda \lambda Y}) \pm 
 \sqrt{(\beta_{YY^2}-\beta_{\lambda \lambda Y})^2 - 4
   \beta_{\lambda\lambda^2}\beta_{\lambda Y^2}} \, , 
\nonumber \\
B_{\pm} &=&  Y(l)\beta_{\pm} - 2
\lambda(l)\beta_{\lambda\lambda^2} \, .
\eea
In addition, the RG running of $\varphi$ is also an essential
ingredient in deriving these formulae [see
Eqs.~(\ref{partition_function_scaling_variable}) and
(\ref{eq:varphi_RG_running})], and it leads to
\beq
\label{eq:RG_for_phi_in_HY_one_loop}
 \varphi (L) = \left [ 1 + Y (l) \beta_{YY^{2}} {\mathrm{log}} \left (
    \frac{\hat{L}}{\hat{l}} \right ) \right
]^{-\frac{2\delta_{Y}}{\beta_{YY^{2}}}} \varphi (l) \, .
\eeq
%

%
%

Results presented in this section can be employed to establish
triviality of the Higgs-Yukawa model, or to search for alternative
scenarios.  When performing the study of the fixed-point structure of
the theory, one can confront the lattice data obtained near a critical
point to Eqs.~(\ref{eq:varphi_RG_running}) and (\ref{recursion_formula}), with the
scaling variable in Eq.~(\ref{full_scaling_variable}).  Our scaling
formulae should fit the data near critical
points that are associated with the Gaussian fixed point.  
The scaling variable contains several unknown parameters, $\lambda
(l)$, $Y(l)$, $M^{2}(l)$, as well as another
constant that accounts for the additive mass renormalisation\footnote{In principle, the integration constant,
  $\varphi (l)$, in Eq.~(\ref{eq:RG_for_phi_in_HY_one_loop}) is not an
  unknown parameter, because it is obtained by matching the lattice
  bare variable to its renormalised counterpart.  This matching
procedure has to be performed, in order to implement the FSS strategy
presented here.  However, as discussed in the next subsection, in
practice there may be unknown parameters associated with it.}.   These
parameters arise as the integration constants in obtaining the
RG running behaviour {\it via} solving
Eq.~(\ref{mass_square_RGE}).
They can in principle be determined from fitting the lattice data.
Since these parameters are also the values of renormalised couplings at
the scale where bare lattice results are matched to a particular
scheme, they are scheme-dependent.   This means that certain choices
of renormalisation scheme may help in reducing the number of free
parameters in the fit.  One such example will be discussed in the next
section.   Finally, we stress that the extension of our analysis in this
section for theories containing more than one fermion doublet can also
be investigated, although it is beyond the scope of our current work.

\section{Numerical test in the O(4) pure scalar model}
\label{sec:O4_numerical_test}
The scaling formulae presented in Sec.~\ref{sec:FSS_RG} are intended 
for the full Higgs-Yukawa model of Eq.~(\ref{eq:HY_lagrangian}), for which they were, in fact, derived
for the first time. Ideally, the formulae and the FSS analysis strategy ought to be directly tested for this case.  
However, this requires a presently prohibitive
numerical effort, and is beyond the scope of this work.   We therefore
decided to perform this test in the simpler, pure scalar O(4) model. 
As explained in the previous section, one
    important ingredient in the derivation
  of the scaling behaviour at one-loop level is the use of 
   Eqs.~(\ref{eq:varphi_RG_running}) and (\ref{recursion_formula}).
This aspect of the procedure is identical in both the Higgs-Yukawa and
the scalar O(4) theories.  The LL-improved mean-field scaling behaviours of these two models
differ only in the detailed logarithmic volume-dependence in the
renormalised couplings.  Therefore, confronting our formulae in the pure
scalar model with lattice simulations is at least directly testing the
validity of Eqs.~(\ref{eq:varphi_RG_running}) and
(\ref{recursion_formula}) for obtaining the FSS properties at this
order.  Furthermore, the scalar O(4) theory can be considered as the 
Higgs-Yukawa model in the limit of vanishing Yukawa coupling.
Since the natural next-stage examination of results presented
in Sec.~\ref{sec:FSS_RG} should be carried out for the Higgs-Yukawa
model with small Yukawa coupling, testing the formulae with
data from lattice simulations in this simpler scalar theory should
lead to non-trivial information regarding the range of parameters and lattice sizes 
these formulae can be applied.
Also on the theoretical side, concentrating on the scalar O(4) model 
simplifies the scaling formulae which significantly eases the analysis of the numerical data.  
\subsection{The scalar O(4) model and its finite-size scaling behaviour}
\label{sec:O4}
The scalar O(4) model is described by the 
pure-scalar part of Eq.~(\ref{eq:HY_lagrangian}).
Using the convention $\boldsymbol{\phi}^T = (\phi_0,\phi_1,\phi_2,\phi_3)$, 
the action of this model with lattice regularisation can be expressed as,
\begin{eqnarray}
 \label{eq:O4_lattice}
 S [ \boldsymbol{\phi} ] &=& \sum_{x \in \Gamma} \l\{ \frac{1}{2}a^2\boldsymbol{\phi}^{T}(x) 
 \l[ \sum_{\mu = 0}^3 \l( \boldsymbol{\phi}({x+a\hat{\mu}}) + \boldsymbol{\phi}({x-a\hat{\mu}}) -2\boldsymbol{\phi}(x) \r) 
   \r] \r\} \nonumber \\
 && \hspace{1.5cm} + \sum_{x \in \Gamma} \l[ a^2M^2_b \boldsymbol{\phi}^{T}(x) \boldsymbol{\phi}(x)  
 +\lambda_b a^4\l( \boldsymbol{\phi}^T(x) \boldsymbol{\phi}(x) \r)^2\r],
\end{eqnarray}
where we use $\Gamma$ to denote the set of lattice sites.  To use information existing in the literature and make educated
guess for values of $\hat{M}_{b}^{2}$ when searching for critical
points in the theory, it can
also be convenient to
express the above action with the change of
variables
\begin{equation}
\label{eq:parameters}
a^2M^2_b  = \frac{1-2\tilde{\lambda}-8\kappa}{\kappa}\;,\;\; \lambda_b = \frac{\tilde{\lambda}}{4\kappa^2}\; .
\end{equation} 
%

The logarithmic dependence of $\hat{L}$ for moments of the
scalar-field zero mode in the scalar O(4) model 
can then be obtained by solving the RGEs, Eq.~(\ref{mass_square_RGE})
at $Y(\rho)=0$,
and integrating the renormalisation 
length scale from $l$ to $L$.
Note that the anomalous dimensions in Eq.~(\ref{mass_square_RGE}) are
universal at the  
leading-order in perturbation theory, although the values of the
renormalised couplings are scheme-dependent.
Applying the scaling formulae, 
the couplings renormalised at $\hat{l}$ are regarded as free
parameters.  They can be determined by fitting the 
FSS behaviour of the considered physical observables.
In this work, we investigate the scaling behaviour of three observables.
The first is the magnetisation, which is defined as the 
expectation value of the scalar zero-mode,
$\langle \hat{\varphi} \rangle$. The second is the
magnetic susceptibility, $\hat{\chi}$, 
\beq
\label{eq:susc_def}
 \hat{\chi} \equiv \hat{V} \l( \langle \hat{\varphi}^2 \rangle - \langle
 \hat{\varphi} \rangle^2 \r) ,
\eeq
where $\hat{V}$ is the lattice four-volume.  Finally, the third is 
the dimensionless four-point function, $\chi^{(4)}$, 
\beq
\label{eq:chi_4_def}
 \hat{\chi}^{(4)} = \chi^{(4)} = \hat{V} \l( 3\langle \hat{\varphi}^2 \rangle^{2} -
 \langle \hat{\varphi}^4 \rangle\r) .
\eeq
These three observables can be
computed in numerical lattice simulations \footnote{In this section, quantities
  with a caret on top are dimensionless in lattice units.  For instance,
  $\hat{\varphi}$ means $a \varphi$.  See the discussion below Eq.~(\ref{eq:ME_renorm}).}.  Their
FSS properties are obtained using 
Eqs.~(\ref{partition_function_scaling_variable}),
(\ref{eq:scaling_var_def}), (\ref{eq:varphibarn_def}) and (\ref{eq:varphi_RG_running}).  Rescaling
them with appropriate powers of the lattice size, $\hat{L}$, it is
straightforward to demonstrate that
\begin{eqnarray}
\label{eq:scaling_mag}
  \langle \hat{\varphi} \rangle \hat{L} &=& A(l)  
  \left [ s\lambda(L) \right ]^{-1/4} \frac{\bar{\varphi}_4(z)}{\bar{\varphi}_3(z)}, \\
\label{eq:scaling_sus}
  \hat{\chi} \hat{L}^{-2}   &=&  A(l)^2 s^{1/2} \left [ \lambda(L)
                                \right ]^{-1/2} 
      \left \{ \frac{\bar{\varphi}_5(z)}{\bar{\varphi}_3(z)} - 
      \left [ \frac{\bar{\varphi}_4(z)}{\bar{\varphi}_3(z)} \right ]^2
                                \right \}, \\
\label{eq:scaling_fourPT}
  \hat{\chi}^{(4)}   &=& A(l)^4\left [ \lambda(L) \right ]^{-1} 
  \left \{ 3\left [ \frac{\bar{\varphi}_5(z)}{\bar{\varphi}_3(z)} \right
                         ]^2 
	- \frac{\bar{\varphi}_7(z)}{\bar{\varphi}_3(z)} \right \} ,
\end{eqnarray}
where $A(l)$ is a free parameter associated with the renormalisation of the scalar
field, and will be discussed in more detail at the end of this subsection.  The one-loop
expression for the quartic coupling renormalised at the length scale
$L$ in this model is
\beq
\label{eq:lambda_L_O4}
\lambda (L) = \frac{\lambda(l)}{1+\beta_{\lambda
    \lambda^2}\lambda(l)\log \l( \frac{\hat{L}}{\hat{l}} \r) } ,
\eeq
where $\beta_{\lambda\lambda^{2}} = 6/\pi^{2}$, as given in Eq.~(\ref{eq:one_loop_coeff_values}).
For the O(4) pure-scalar field theory, the LL corrections in
$M^{2}$ and $\lambda^{-1/2}$ cancel.  This means that at one-loop order the combination 
$M^{2}\lambda^{-1/2}$ is independent of the renormalisation
scale, although its value is scheme-dependent.  For
this reason, the scaling variable, $z$, at this order in this theory is
\beq
 \label{eq:O4_scaling_variable}
 z = \sqrt{s} M^2 (l) L^2 \lambda^{-1/2}(l)  = \sqrt{s} \hat{M}^2 (l) \hat{L}^2 \lambda^{-1/2}(l).
\eeq
That is, it does not feature any logarithmic $L{-}$dependence.

\begin{figure}
\begin{center}
\includegraphics[width=7cm, height=6cm]{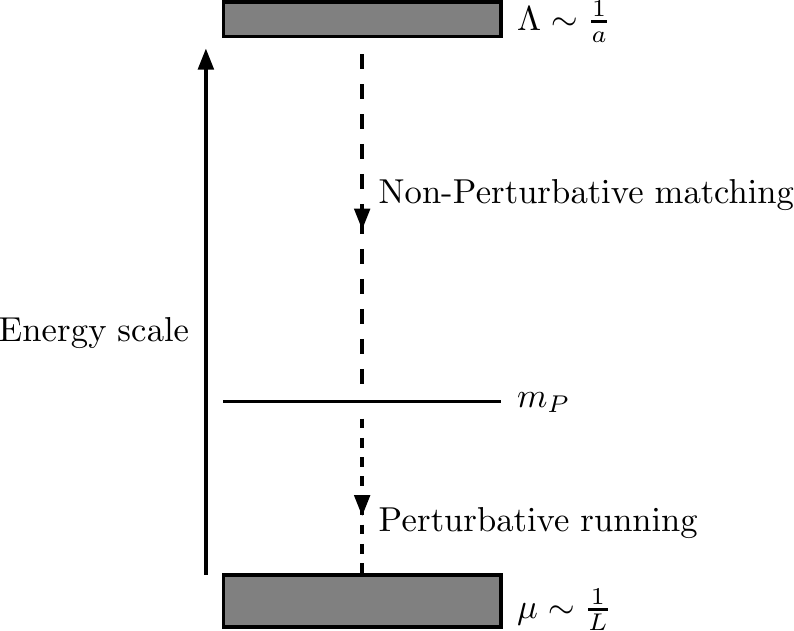}
 \caption[Scales in our strategy]{
    The scales involved in our FSS strategy with the
    on-shell renormalisation scheme.
    Starting from the lattice theory at scale $\Lambda \propto 1/a$, 
    we non-perturbatively match bare quantities to the on-shell
    renormalisation scheme at $m_P$, from where 
    we perform a leading-order perturbative running to the energy scale 
    $\mu \sim 1/L$. 
    \label{fig:scale_hierarchy}
}
\end{center}
\end{figure}

The FSS formulae, Eqs.~(\ref{eq:scaling_mag}),
(\ref{eq:scaling_sus}) and (\ref{eq:scaling_fourPT}) contain four
free parameters, $A(l)$, $\lambda(l)$, $M(l)$ and an
additive renormalisation constant in the relationship between the bare
quadratic coupling. $M^{2}_{b}$ in Eq.~(\ref{eq:O4_lattice}), and $M^{2}(l)$.   Given that the
scaling functions, $\bar{\varphi}_{3,4,5,7}(z)$, are already quite
complex, see Eq.~(\ref{recursion_formula}), it is challenging to
determine all four parameters {\it via} fitting these functions
to the lattice data.  On the other hand, since these free parameters
are related to properties of renormalisation of the theory, one may be
able to simplify the numerical analysis procedure by working with a
particular scheme.  For this purpose, in this project we implement the
FSS strategy using the on-shell renormalisation scheme, which is
defined by relating a pole mass of the scalar field in the
infinite-volume limit, denoted as
$m_{P}$, to the renormalised coupling $M(l)$ at the scale
\beq
\label{eq:on_shell_scheme_scale_identification}
l = \frac{1}{m_{P}} .
\eeq
This pole mass, $m_P$, must be much smaller than the cutoff scale such that
the running from $\hat{m}_P^{-1}$ to $\hat{L}$ can be well
described by leading-order perturbation theory.   Also, since $\hat{m}_{P}$
will in practice  be computed by studying lattice numerical data for the
scalar-field propagators, one has to ensure that its infinite-volume
extrapolation is under control.  
The above conditions lead to the requirements that $m_PL \gtrsim 1$ as well as $\hat{m}_P \ll 1$.
This hierarchy of scales and the FSS strategy we will employ are depicted in
Fig.~\ref{fig:scale_hierarchy}.   

In the symmetric phase where all
four components of the scalar field are equivalent, $m_{P}$ can be
extracted by studying the propagator of one of $\phi_{0,1,2,3}$, and
it is connected to the renormalised $M$ through the simple relation in
the on-shell scheme,
\beq
\label{eq:m_P_to_M_symmetric_phase}
m_P^2 = M^2 (m_{P}^{-1}) \mbox{ }\mbox{ }\mbox{ }\mbox{ }\text{(symmetric phase)},
\eeq
which is indicated by the effective potential of the theory.   In the
broken phase, $m_{P}$ is identified to be the pole mass of the ``Higgs
mode'' that is the field variable obtained by performing a projection operation,
\beq
\label{eq:Higgs_mode_projection}
 H (x) = \sum_{\alpha=0}^{3} \frac{\varphi_{\alpha}}{\varphi}
 \phi_{\alpha} (x) \, , \mbox{ }\forall x \in \Gamma ,
\eeq
where $\varphi_{\alpha}$ and $\varphi$ are defined in Eqs.~(\ref{eq:varphi_def})
and (\ref{eq:mod_varphi_def}), respectively.   This Higgs field is
massive in this phase.   One can also construct
the three ``Goldstone modes'' that are perpendicular to $H(x)$ in the
internal O(4) space.  These Goldstone modes are denoted as $G(x)$ in
this work, and they should be massless.  It is straightforward to show
that in the broken phase, the implementation of the on-shell scheme
results in the condition
\beq
\label{eq:m_P_to_M_broken_phase}
m_P^2 = -2 M^2 (m_{P}^{-1}) \mbox{ }\mbox{ }\mbox{ }\mbox{ }\text{(broken phase)}.
\eeq

From the above discussion, it is obvious that in the
numerical analysis for the scaling test, use of the on-shell
renormalisation scheme reduces the number of free parameters from four
to two.  Since the renormalised mass can be obtained
non-perturbatively from studying
the lattice data for the scalar-field propagators, one only needs to determine $A(m_{P}^{-1})$
and $\lambda(m_{P}^{-1})$ from the fits to Eqs.~(\ref{eq:scaling_mag}),
(\ref{eq:scaling_sus}) and (\ref{eq:scaling_fourPT}).  One aspect in the non-trivial
verification of our FSS formulae is the demonstration
that employing three different quantities ($\la\hat{\varphi}\ra$,
$\hat{\chi}$ and $\hat{\chi}^{(4)}$) to extract $A(m_{P}^{-1})$
and $\lambda(m_{P}^{-1})$ leads to compatible results.

We close the discussion of the FSS analysis strategy for the O(4) scalar model by
elaborating on the introduction of the parameter
$A(l)=A(m_{P}^{-1})$ in Eqs.~(\ref{eq:scaling_mag}),
(\ref{eq:scaling_sus}) and (\ref{eq:scaling_fourPT}).   From
Eq.~(\ref{eq:varphi_RG_running}), it can be seen that there is no
integration constant for the running of the moments of the
renormalised scalar-field zero mode, if the zero mode is matched from the
lattice-regularised theory to the chosen renormalisation scheme, the
latter being the on-shell scheme in this work.   In general this
matching has to be carried out non-perturbatively, since couplings at
high-energy scale can be strong in this theory.   
Very frequently, it is
much more convenient to first match the field variable to an
``intermediate renormalisation scheme'', {\it
  e.g.}, a momentum-subtraction (MOM) scheme.  The parameter
$A(m_{P}^{-1})$ is incorporated in the scaling formulae to
account for the connection between the intermediate and the on-shell schemes.

\subsection{Simulation details}
\label{sec:simulate_O4}
%

To implement the strategy for testing our FSS formulae as presented in
the last subsection, we perform
lattice simulations for the O(4) pure scalar field theory described by
the action in Eq.~(\ref{eq:O4_lattice}).  In order to realise the
hierarchy of scales,
\beq
\label{eq:scale_hierarchy}
 a \ll m_{P}^{-1} \lesssim L ,
\eeq
lattices with large size are necessary.  For this purpose, we carry
out calculations with the spatial volume $L^{3}$ at
\beq
\label{eq:sim_latt_size}
 \hat{L} = L/a = 36, 40, 44, 48, 56,
\eeq
and the temporal extent
\beq
\label{eq:latt_temp_asymm}
 \hat{L}_{t} = s \hat{L}\, , \mbox{ }\mbox{ } s = 2 .
\eeq
Furthermore, since our goal is to test the scaling properties
governed by the Gaussian fixed point, we choose to proceed with a weak
bare quartic coupling,
\beq
\label{eq:sim_lambda_b}
 \lambda_{b} = 0.15 ,
\eeq
and then simulate at various values of the bare quadratic coupling,
$\hat{M}_{b}^{2}$, to perform a scan and search for phase transitions.  
Through the initial study of the magnetisation and the susceptibility,
we then identify six choices of $\hat{M}_{b}^{2}$ that are close to
the critical point.  They are
\begin{itemize}
 \item $\hat{M}_{b}^{2} = -0.54141$, $-0.54189$ and $-0.54236$,
   corresponding to $\kappa = 0.131300$, $0.131308$ and $0.131316$.
   These points are in the symmetric phase.
 \item $\hat{M}_{b}^{2} = -0.54334$, $-0.54349$ and $-0.54378$,
   corresponding to $\kappa = 0.1313325$, $0.131335$ and $0.131340$.
   These points are in the broken phase.
\end{itemize}
Simulations at other $\kappa$  values have also been executed.
However, these other data points are too far away from the
critical point, which makes
it impractical to include them in the scaling test.

Since we are carrying out lattice computations near a critical
point in the theory, it is necessary to employ an efficient algorithm
that can reduce the effects of critical slowing down.   For this, we
implement a method that combines the Metropolis and the
cluster~\cite{Wolff:1988uh} updates.   A complete sweep in this
algorithm consists of one Metropolis step, where 
all even (odd) lattice sites are updated sequentially,  
and $32$ cluster updates \cite{Wolff:1988uh}, in each of which at
least $10\%$ of the lattice sites are included in the clusters.   For
each choice of $\hat{M}_{b}^{2}$, one to three
Markov chains are created with hot start.  In every Markov chain, the first 
200 sweeps are discarded for thermalisation purpose.   We then proceed
to perform measurements on about 20,000 scalar configurations for each set of bare
parameters, at the interval of three sweeps.

\subsection{Analysis and numerical results}
\label{sec:scaling_test}
The typical autocorrelation time for the data points used in our
analysis is about 20 configurations.  In view of this, we bin the data with
the bin size being 40 configurations.  Statistical analysis is carried
out by creating 2,500 bootstrap samples for each bare-parameter set.

\begin{figure}
\begin{center}
\includegraphics[width=0.45\columnwidth]{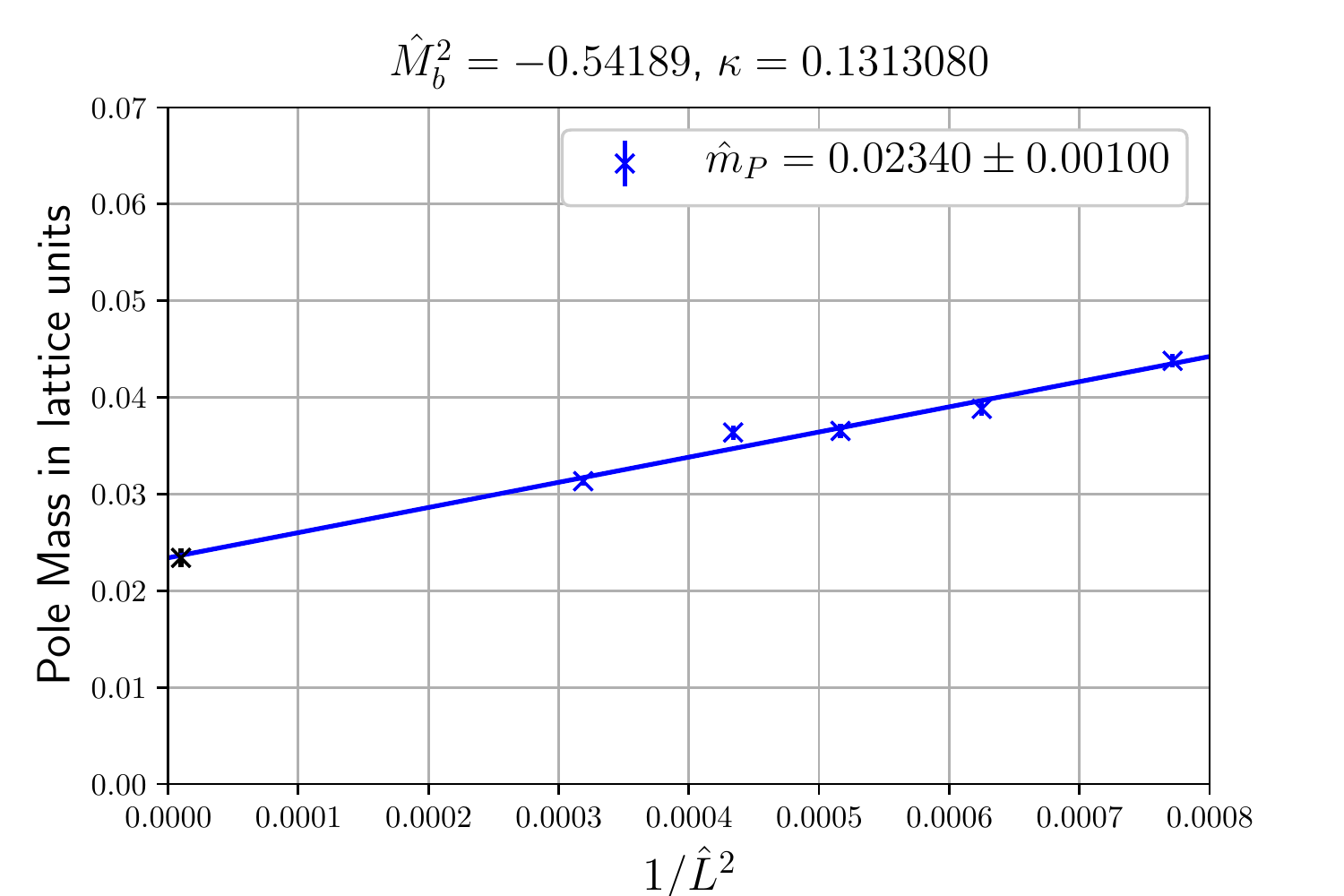}
\includegraphics[width=0.45\columnwidth]{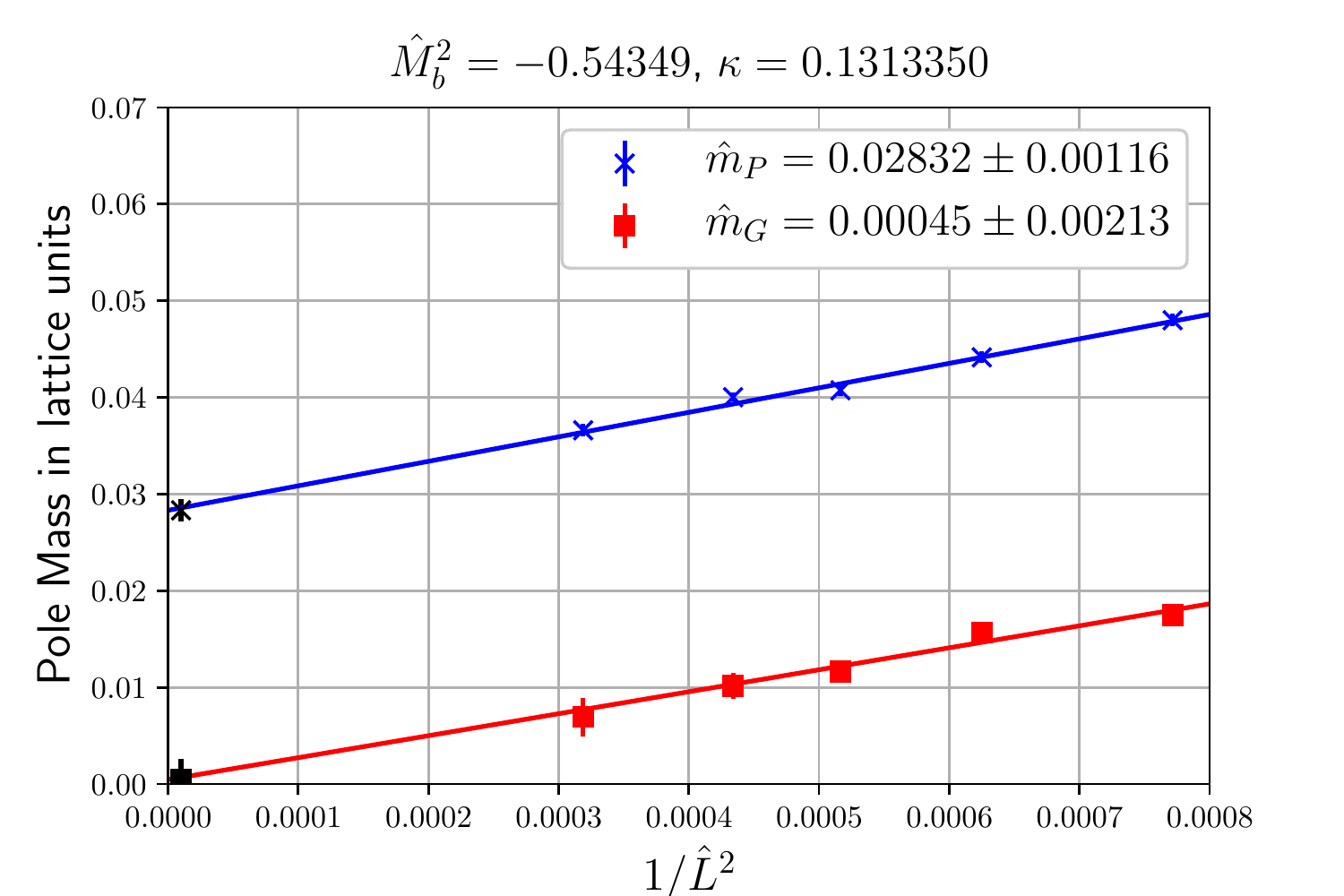}
\hspace*{0cm}(a)\hspace*{7.80cm}(b)
  \caption{Examples of the infinite-volume extrapolation for the pole
    mass in the symmetric (a) and the broken (b) phases using the
    ansatz of Eq.~(\ref{eq:infinitepole}).  Results for the
    Goldstone-boson mass, computed from the average for the
    propagators of the three Goldstone modes, in the broken phase are also shown in (b).
  }
\label{fig:m_P_inf_vol_extrap}
\end{center}
\end{figure}

\begin{figure}
\begin{center}
\includegraphics[width=16cm,height=11cm]{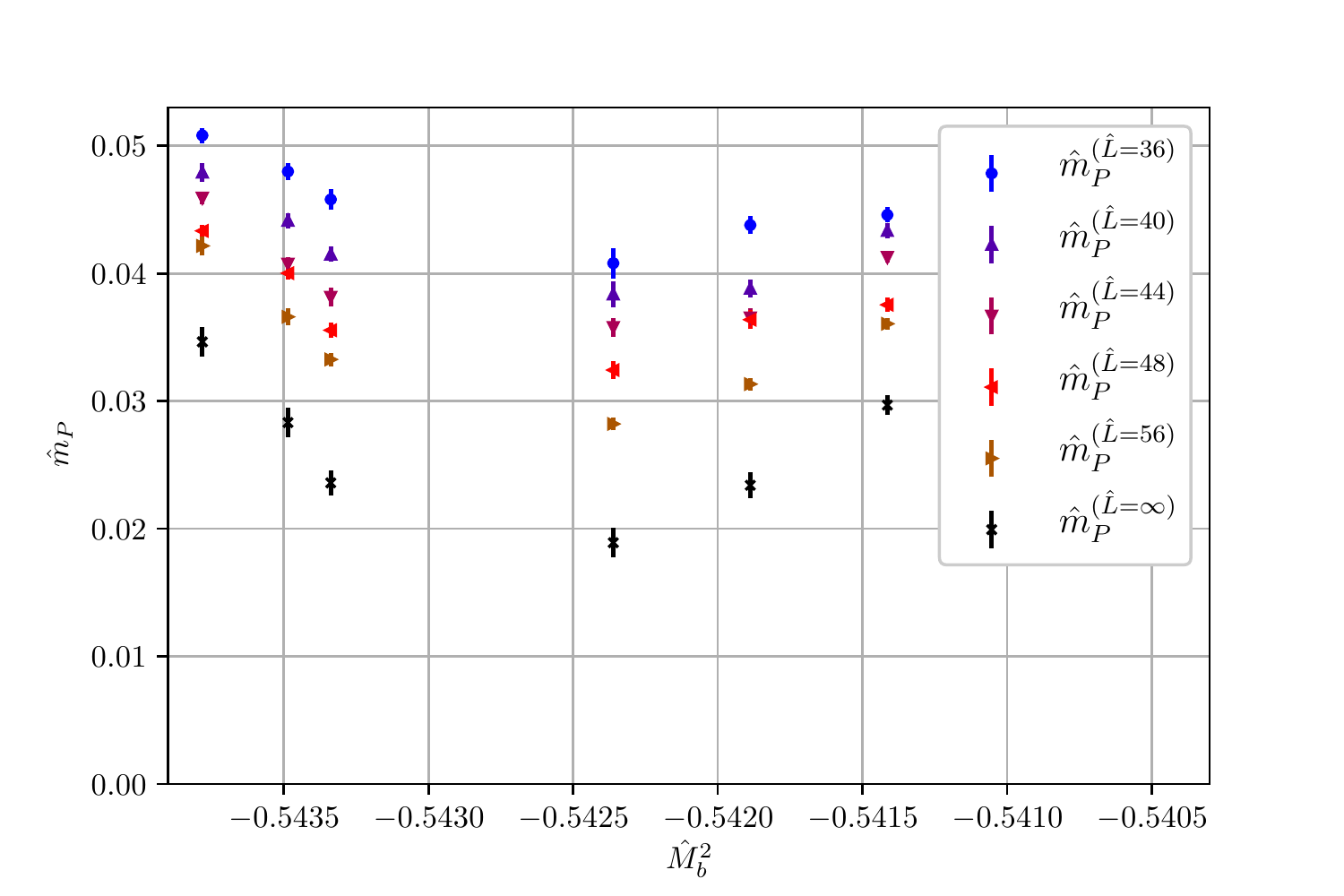}
 \caption[Infinite volume extrapolations of masses]{
    The infinite volume extrapolations of the pole mass $m_P$ in lattice units.  
}
\label{fig:IVextra}
\end{center}
\end{figure}

In our strategy for the FSS test explained in Sec.~\ref{sec:O4},  the
pole mass in the infinite-volume limit, $m_{P}$, is an essential
ingredient.  The pole mass is extracted by studying the relevant
scalar-field propagator in momentum space using our
numerical data\footnote{In the symmetric
  phase, we average over the modes of $\phi_{0,1,2,3}$.}, and then extrapolated to the infinite-volume limit at fixed
lattice spacing, {\it i.e.}, at fixed value of $\kappa$.   This
extrapolation is performed using an ansatz inspired by the results in Ref.~\cite{Hasenfratz:1989pk}, 
\begin{equation}
 \hat{m}^{(\hat{L})}_P = \hat{m}^{(L = \infty)}_P + \frac{{\mathcal{A}}}{\hat{L}^{2}}\; ,
\label{eq:infinitepole}
\end{equation}
where $\hat{m}^{(\hat{L})}_P$ denotes the pole mass measured from our
lattice data in finite volume, while $\hat{m}^{(L = \infty)}_P$ and
${\mathcal{A}}$ are fit parameters, with $\hat{m}^{(L = \infty)}_P$
being the infinite-volume result, $\hat{m}_{P}$.   We have also used the same
ansatz to investigate the mass of the Goldstone bosons, and find that
it is always compatible with zero in the infinite-volume limit.
Figure~\ref{fig:m_P_inf_vol_extrap} shows examples of this
extrapolation at two values of the bare coupling $\hat{M}_{b}^{2}$.
Results of the pole mass measured on all lattice volumes and extrapolated to
the $\hat{L} \rightarrow \infty$ limit are summarised in
Table~\ref{tab:m_P} and Fig.~\ref{fig:IVextra}.
\begin{table}
 \centering
 \begin{tabular}{ c | c | c | c | c | c | r }
 & \multicolumn{3}{ |c|  }{ Symmetric phase } & \multicolumn{3}{|c}{Broken phase} \\
 \hline
  $\hat{M}^2_b$  & $-0.54141$    & $-0.54189$    & $-0.54236$ & $-0.54334$     & $-0.54349$     & $-0.54378$ \\ 
  $\kappa$    & $0.131300$    & $0.131308$    & $0.131316$    & $0.1313325$    & $0.131335$     & $0.131340$ \\ \hline
  $\hat{L} = 36$    & $0.04459(59)$ & $0.04378(71)$ & $0.04078(119)$ &  $0.04580 (83)$ & $0.04800 (65)$ & $0.05082 (60)$\\ \hline 
  $\hat{L}= 40$    & $0.04334(62)$ & $0.03882(72)$ & $0.03840(102)$ &  $0.04151 (59)$ & $0.04415 (59)$ & $0.04794 (72)$ \\ \hline
  $\hat{L} = 44$    & $0.04129(45)$ & $0.03652(77)$ & $0.03579(74)$ &  $0.03813 (75)$ & $0.04076 (57)$ & $0.04590 (51)$ \\ \hline
  $\hat{L} = 48$    & $0.03755(55)$ & $0.03638(73)$ & $0.03244(69)$ &  $0.03554 (62)$ & $0.04002 (47)$ & $0.04333 (50)$ \\ \hline
  $\hat{L} = 56$    & $0.03606(44)$ & $0.03134(50)$ & $0.02823(52)$ &  $0.03325 (52)$ & $0.03658 (67)$ & $0.04216 (76)$ \\ \hline
  $\hat{L} = \infty$& $0.02959(79)$ & $0.02340(100)$ & $0.01890(115)$ &  $0.02359(100)$ & $0.02831(116)$ & $0.03464(114)$
 \end{tabular}
  \caption[Infinite-volume extrapolation]{
 Results of the pole mass in lattice units, $\hat{m}_{P}$, measured on
 our lattice, and its extrapolation to the infinite-volume limit.
  \label{tab:m_P}
  }
\end{table}

When examining the scalar-field propagators, we observe that the
residue for all the propagators computed from the lattice data is consistent with one.  This means
that within our numerical precision, the field variables are already
naturally ``matched'' to a MOM scheme.  As discussed at the end of
Sec.~\ref{sec:O4}, the connection between this MOM scheme and the
on-shell renormalisation scheme can be accounted for by introducing a
fit parameter [$A(m_{P}^{-1})$ in  Eqs.~(\ref{eq:scaling_mag}), (\ref{eq:scaling_sus}) and
(\ref{eq:scaling_fourPT})] in the analysis strategy.  

Once the infinite-volume pole mass in lattice units, $\hat{m}_{P}$, is
determined for each choice of $\hat{M}_{b}^{2}$, it can be used to
obtain the corresponding renormalised quadratic coupling,
$\hat{M}^{2}(m_{P}^{-1})$, through
Eqs.~(\ref{eq:m_P_to_M_symmetric_phase}) and
(\ref{eq:m_P_to_M_broken_phase}).   This quadratic coupling is an
important ingredient in constructing the scaling variable, $z$, in
Eq.~(\ref{eq:O4_scaling_variable}).  Since its value at each lattice
spacing can be determined numerically from the above procedure, the
remaining fit parameters in the scaling formulae of
Eqs.~(\ref{eq:scaling_mag}), (\ref{eq:scaling_sus}) and
(\ref{eq:scaling_fourPT}) are $A(m_{P}^{-1})$ and
$\lambda(m_{P}^{-1})$.  From the discussion in Sec.~\ref{sec:O4}, it
is clear that this infinite-volume pole mass need not take the same
value in the symmetric and the broken phases.  For this reason, we
distinguish $A(m_{P}^{-1})$ and $\lambda(m_{P}^{-1})$ in these two
phases, and denote them as $(A_{sy}, \lambda_{sy})$ and $(A_{br},
\lambda_{br})$, respectively.  That is, in fitting our data for the
magnetisation, the susceptibility and the fourth moment of the
scalar-field zero mode to Eqs.~(\ref{eq:scaling_mag}),
(\ref{eq:scaling_sus}) and (\ref{eq:scaling_fourPT}), there are four
unknown parameters ($A_{sy}$, $\lambda_{sy}$, $A_{br}$ and $\lambda_{br
}$) to be determined.   With the knowledge of these four parameters,
the scaling formulae then predict that the ``rescaled'' quantities,
\begin{eqnarray}
\label{eq:rescaled_mag}
 \langle \hat{\varphi} \rangle_{rs} &=& \langle \hat{\varphi} \rangle \hat{L} 
 \times A(m_P^{-1})^{-1}\left [ s\lambda(L) \right ]^{1/4}, \\
\label{eq:rescaled_sus}
 \hat{\chi}_{rs}  &=& \hat{\chi} \hat{L}^{-2} \times A(m_P^{-1})^{-2}
                      s^{-1/2} \left [ \lambda(L) \right ]^{1/2}, \\
\label{eq:rescaled_fourPT}
 \hat{\chi}^{(4)}_{rs} &=& \hat{\chi}^{(4)} \times
                           A(m_P^{-1})^{-4}\left [ \lambda(L) \right
                           ],
\end{eqnarray}
with $\lambda (L)$ given in Eq.~(\ref{eq:lambda_L_O4}) at $l=1/m_{P}$,
exhibit universality.  That is, when plotted agains the
scaling variable, $z$, each of them should collapse to a common curve.
We stress again that in the above equations, $[A(m_P^{-1}),
\lambda(m_{P}^{-1})]$ actually means $(A_{sy},\lambda_{sy})$ and
$(A_{br},\lambda_{br})$ in the symmetric and broken phases, respectively.

\begin{figure}
\begin{center}
\includegraphics[width=0.45\columnwidth]{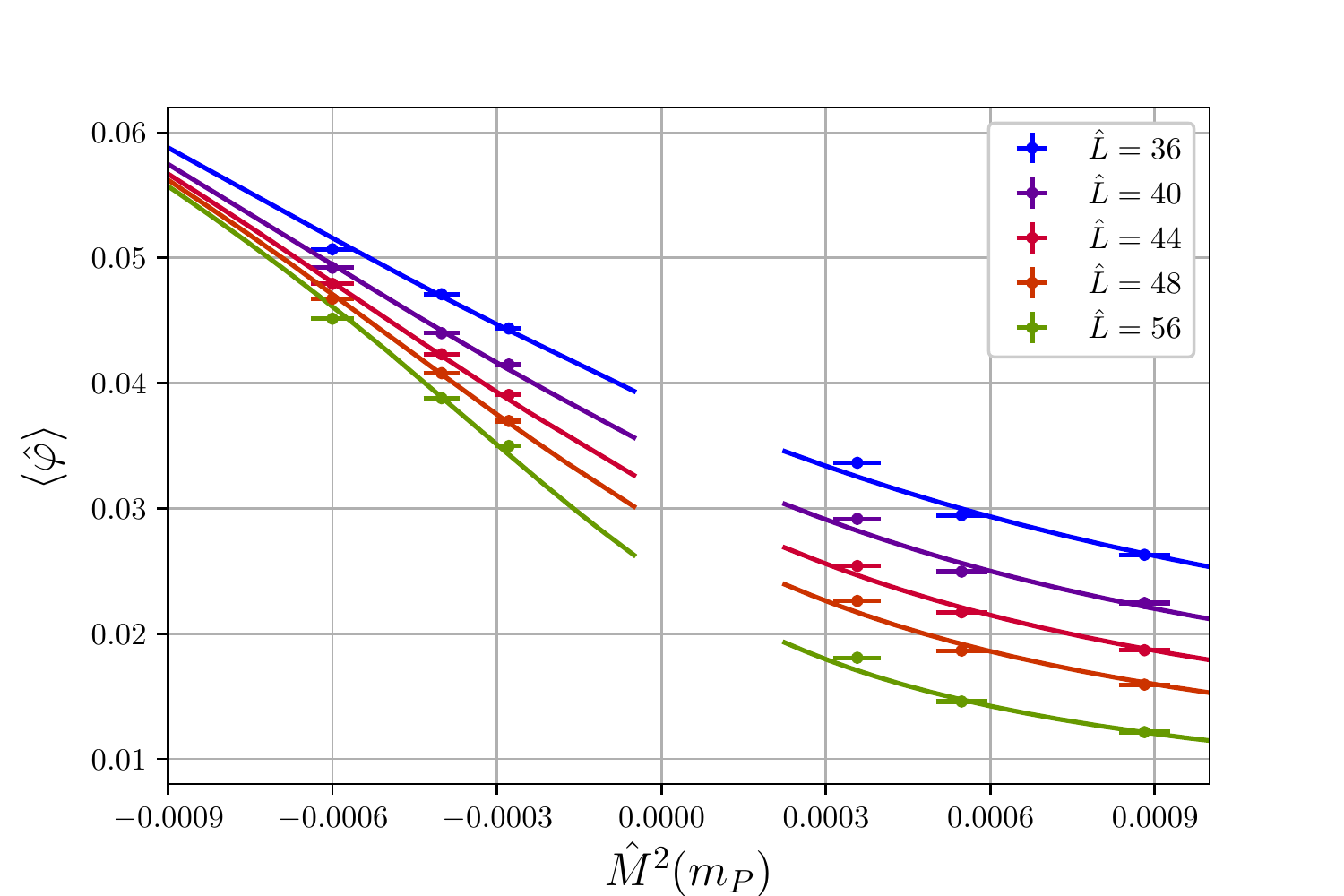}
\includegraphics[width=0.45\columnwidth]{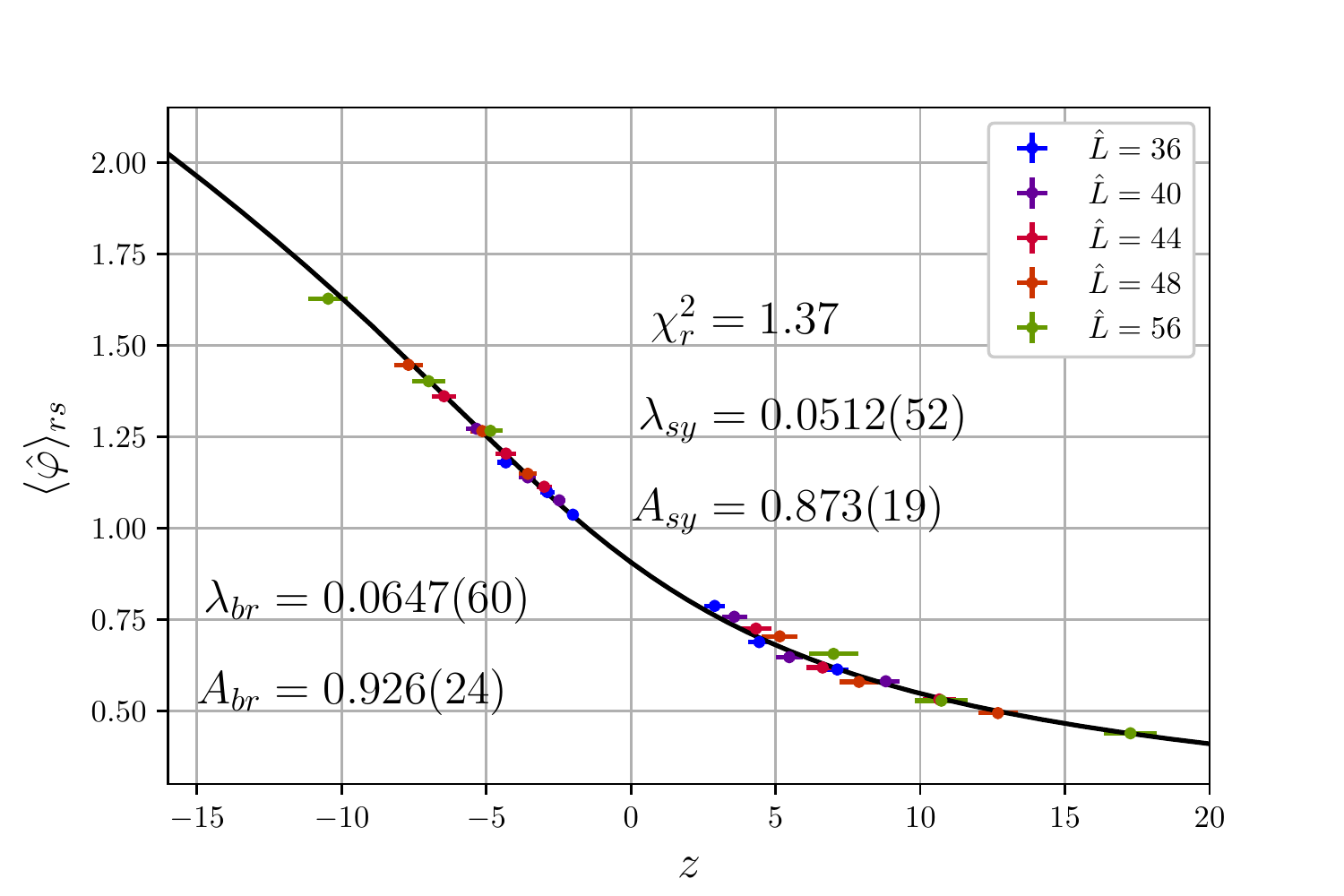}
\hspace*{0cm}(a)\hspace*{7.80cm}(b)\\
\includegraphics[width=0.45\columnwidth]{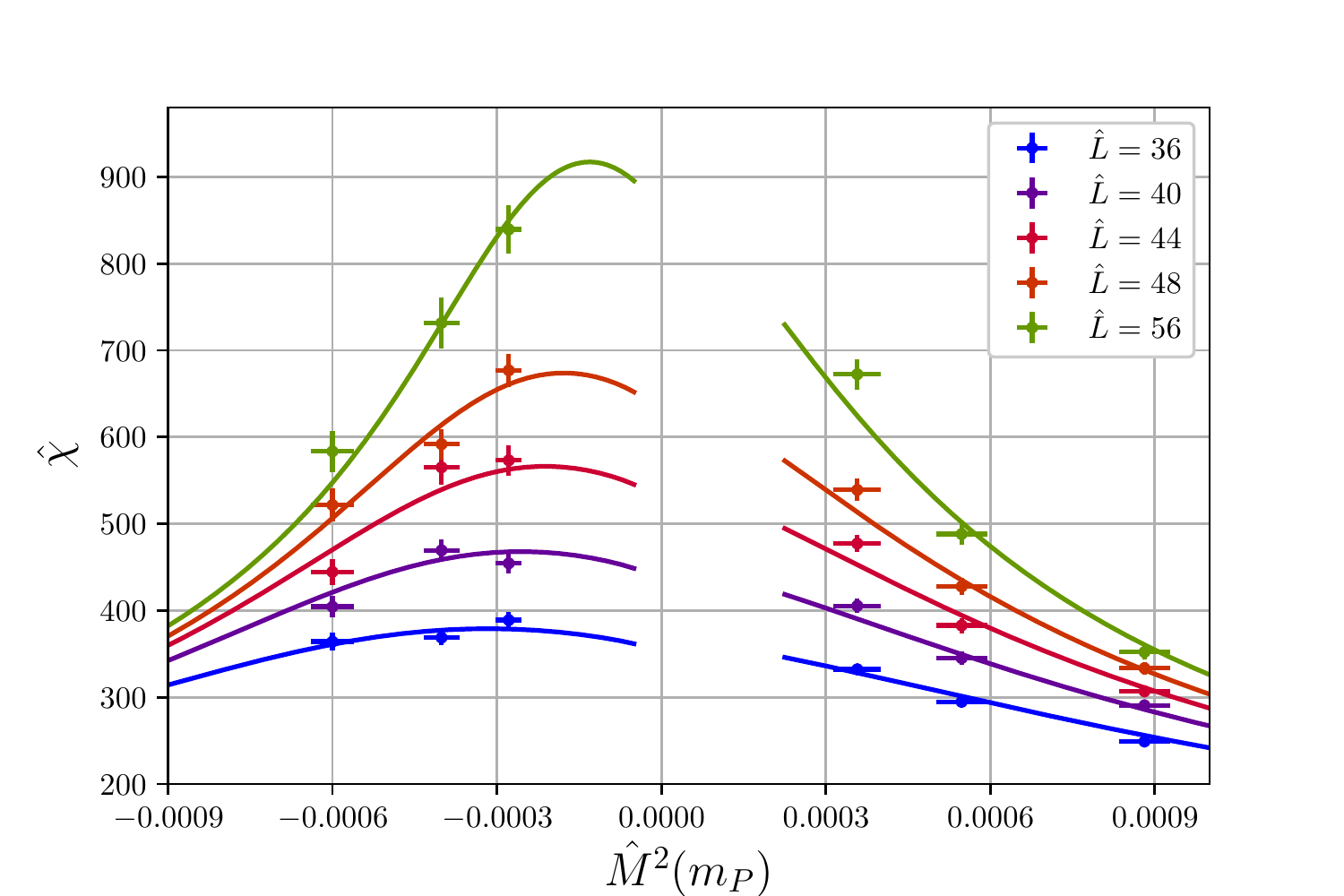}
\includegraphics[width=0.45\columnwidth]{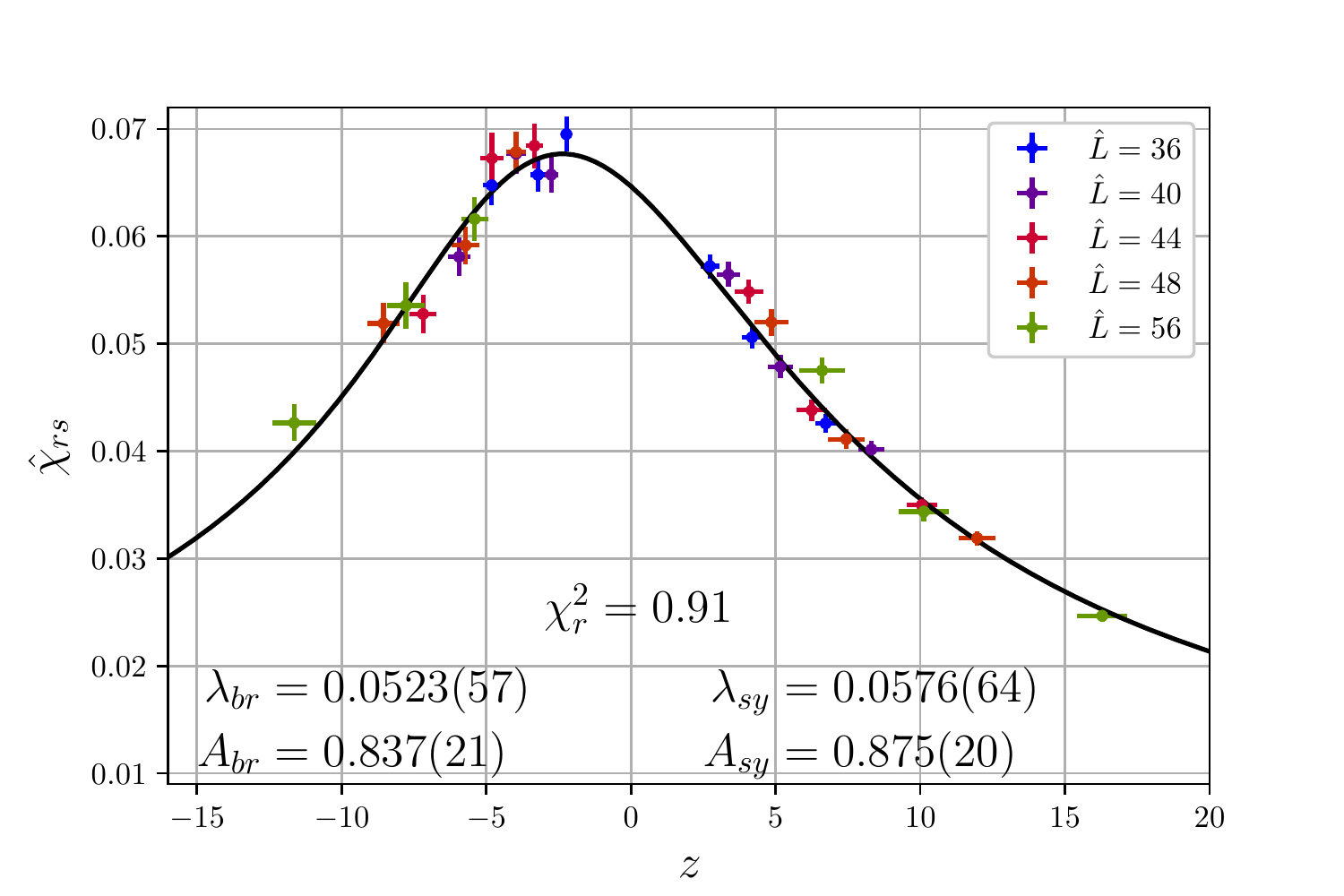}
\hspace*{0cm}(c)\hspace*{7.80cm}(d)\\
\includegraphics[width=0.45\columnwidth]{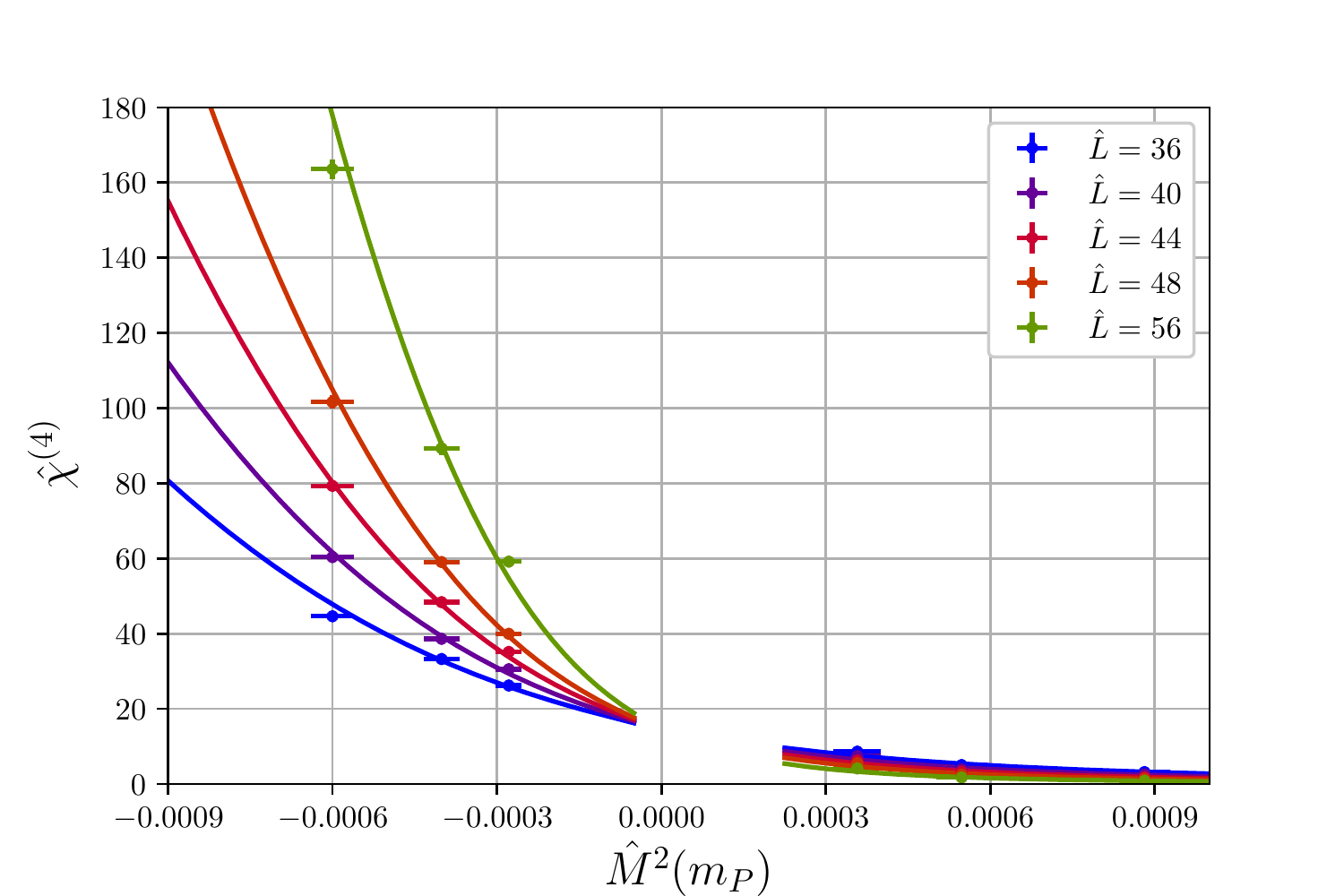}
\includegraphics[width=0.45\columnwidth]{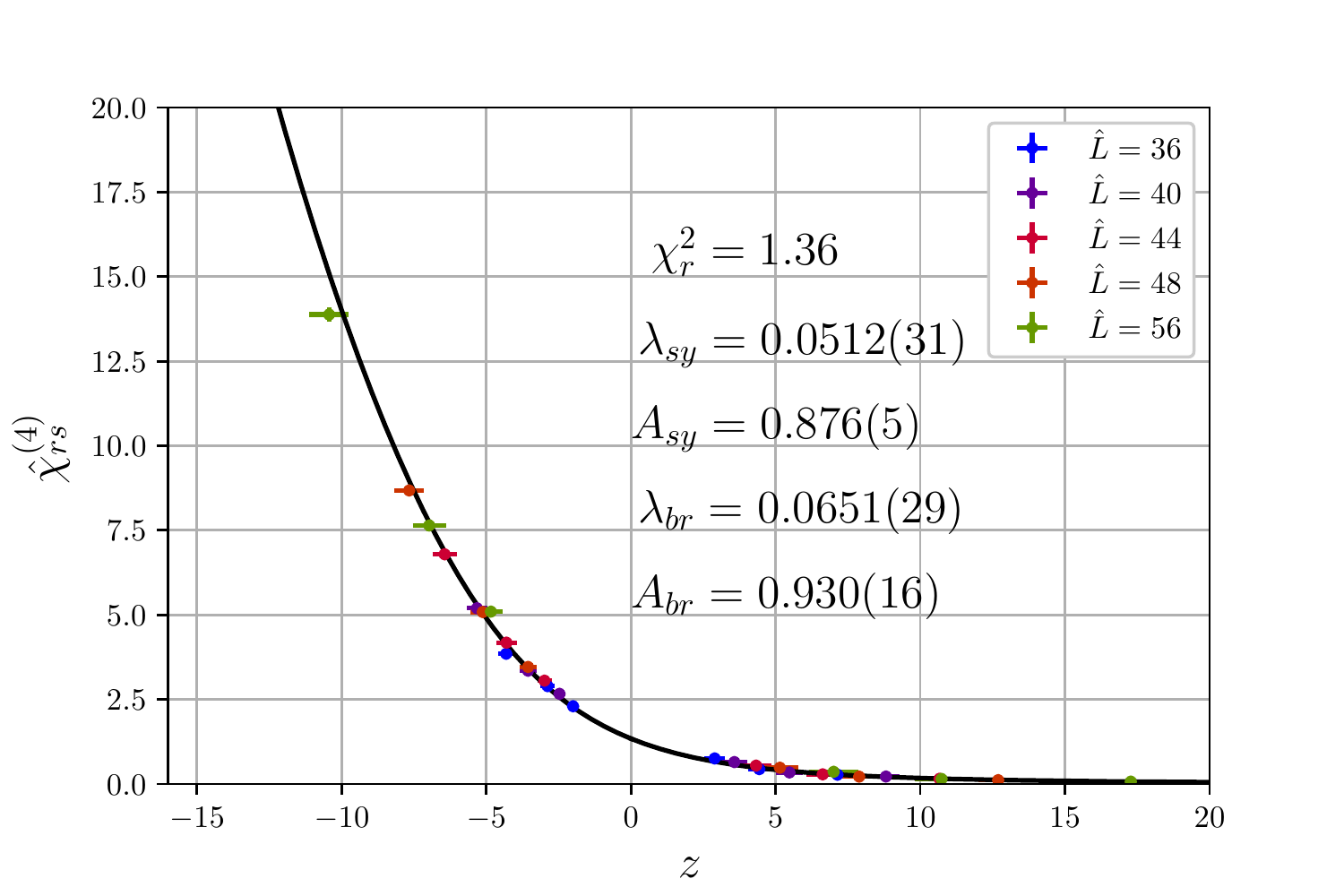}
\hspace*{0cm}(e)\hspace*{7.80cm}(f)\\
  \caption{
Results of the scaling tests {\it via } fitting $\la \hat{\varphi} \ra$, $\hat{\chi}$
and $\hat{\chi}^{(4)}$ separately to Eqs.~(\ref{eq:scaling_mag}),
(\ref{eq:scaling_sus}) and (\ref{eq:scaling_fourPT}) are displayed in
(a), (c) and (e).  The corresponding rescaled quantities defined in Eqs.~(\ref{eq:rescaled_mag}),
(\ref{eq:rescaled_sus}) and (\ref{eq:rescaled_fourPT}) are plotted
against the scaling variable in (b), (d) and (f).
  }
\label{fig:Individual_fits}
\end{center}
\end{figure}

Results of the individual fits to the FSS formulae, Eqs.~(\ref{eq:scaling_mag}),
(\ref{eq:scaling_sus}) and (\ref{eq:scaling_fourPT}), for the magnetisation, the
susceptibility and the fourth zero-mode moment, are shown in 
in Figs.~\ref{fig:Individual_fits}(a),(c),(e).  
Their corresponding
rescaled quantities defined in Eqs.~(\ref{eq:rescaled_mag}),
(\ref{eq:rescaled_sus}) and (\ref{eq:rescaled_fourPT}) are plotted
against the scaling variable
in Figs.~\ref{fig:Individual_fits}(b),(d),(f).
As can be seen, these fits lead to good $\chi^{2}$ per degree of
freedom (denoted as $\chi^{2}_{r}$ in this work), and scaling behaviour is indeed observed.   More importantly,
these separate fits for $\la \hat{\varphi} \ra$, $\hat{\chi}$ and
$\hat{\chi}^{(4)}$ result in compatible values of $A_{sy}$,
$\lambda_{sy}$, $A_{br}$ and $\lambda_{br}$.  This demonstrates
evidence for the validity of the strategy we designed
in Sec.~\ref{sec:FSS_RG}.

\begin{figure}
\begin{center}
\includegraphics[width=0.45\columnwidth]{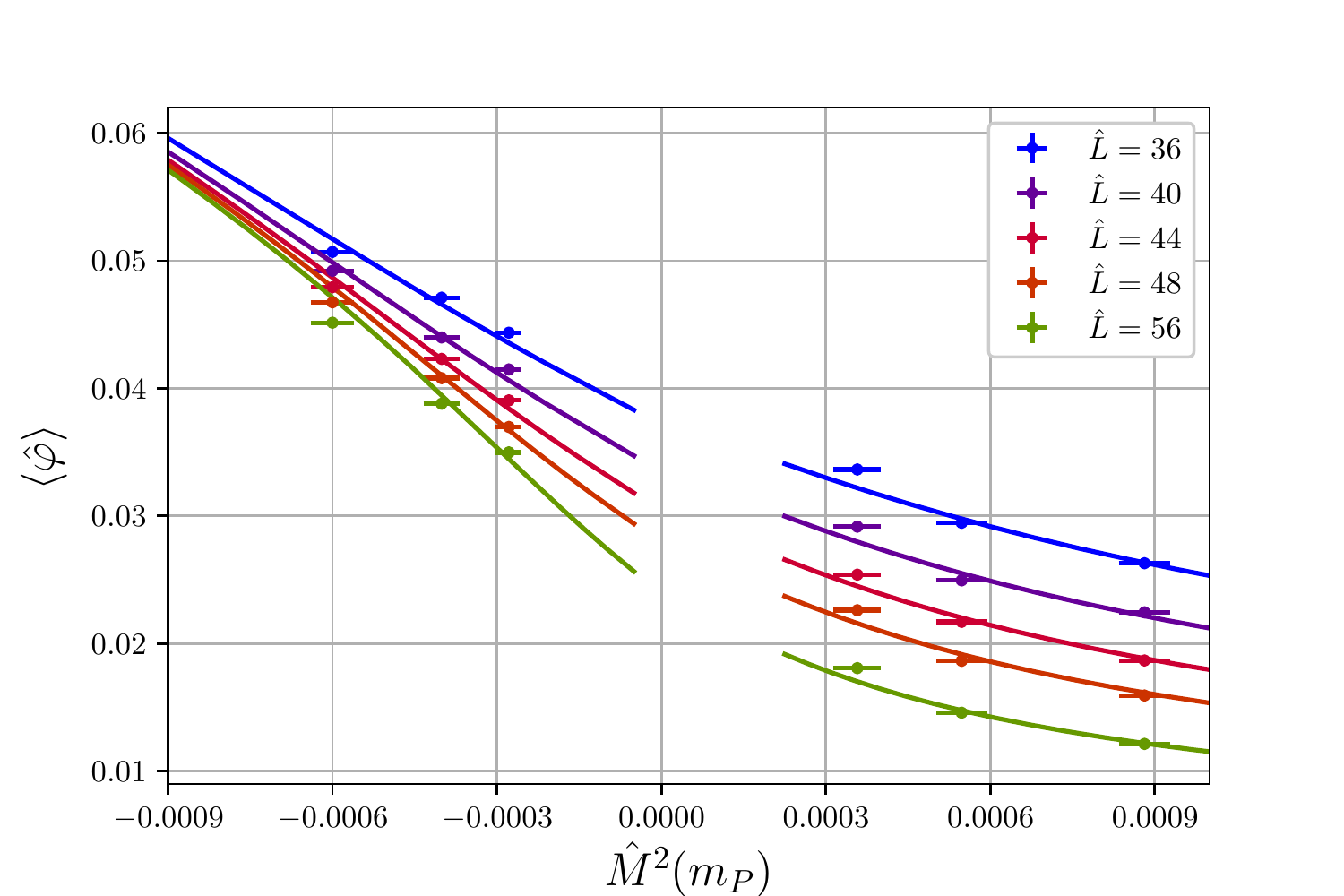}
\includegraphics[width=0.45\columnwidth]{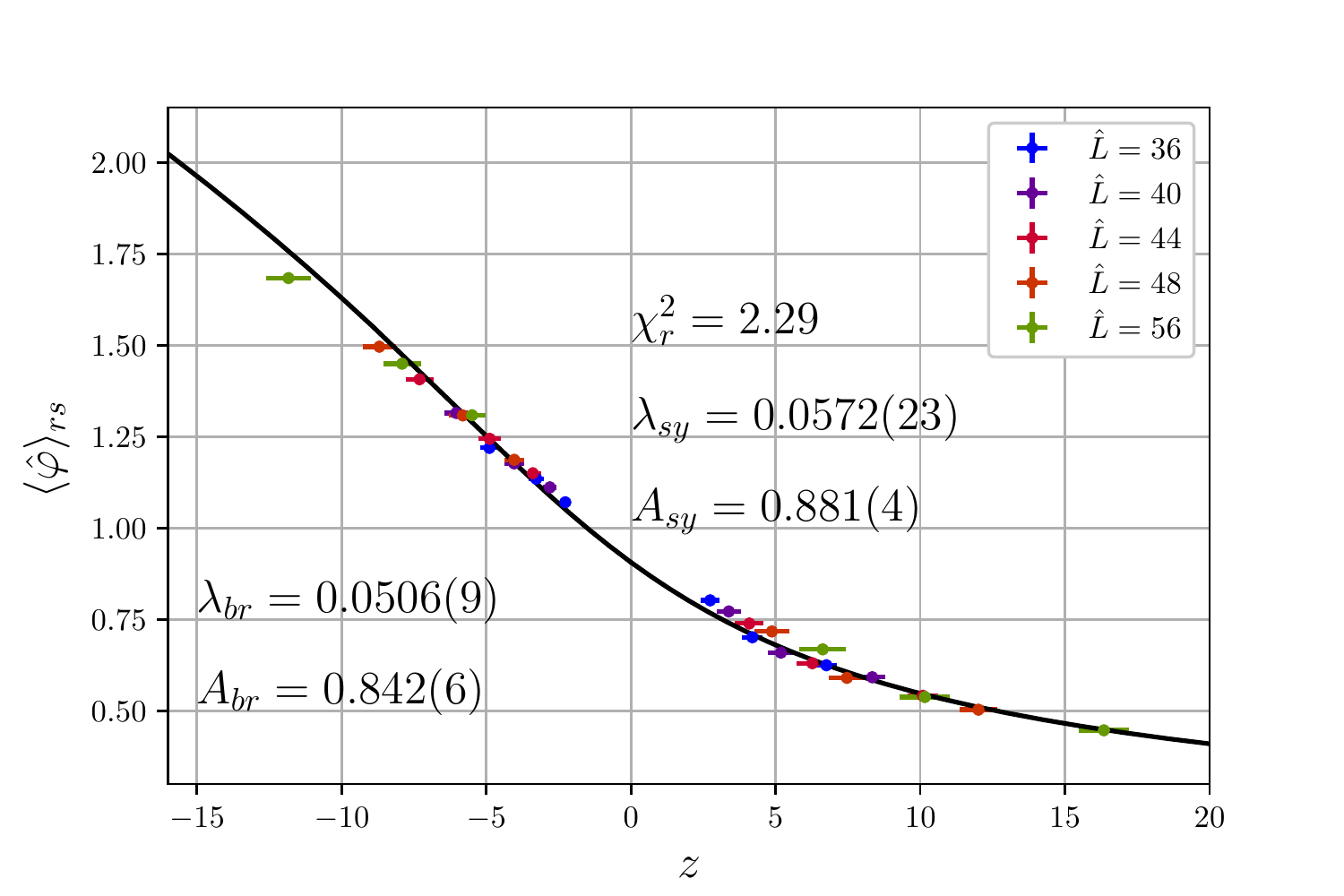}
\hspace*{0cm}(a)\hspace*{7.80cm}(b)\\
\includegraphics[width=0.45\columnwidth]{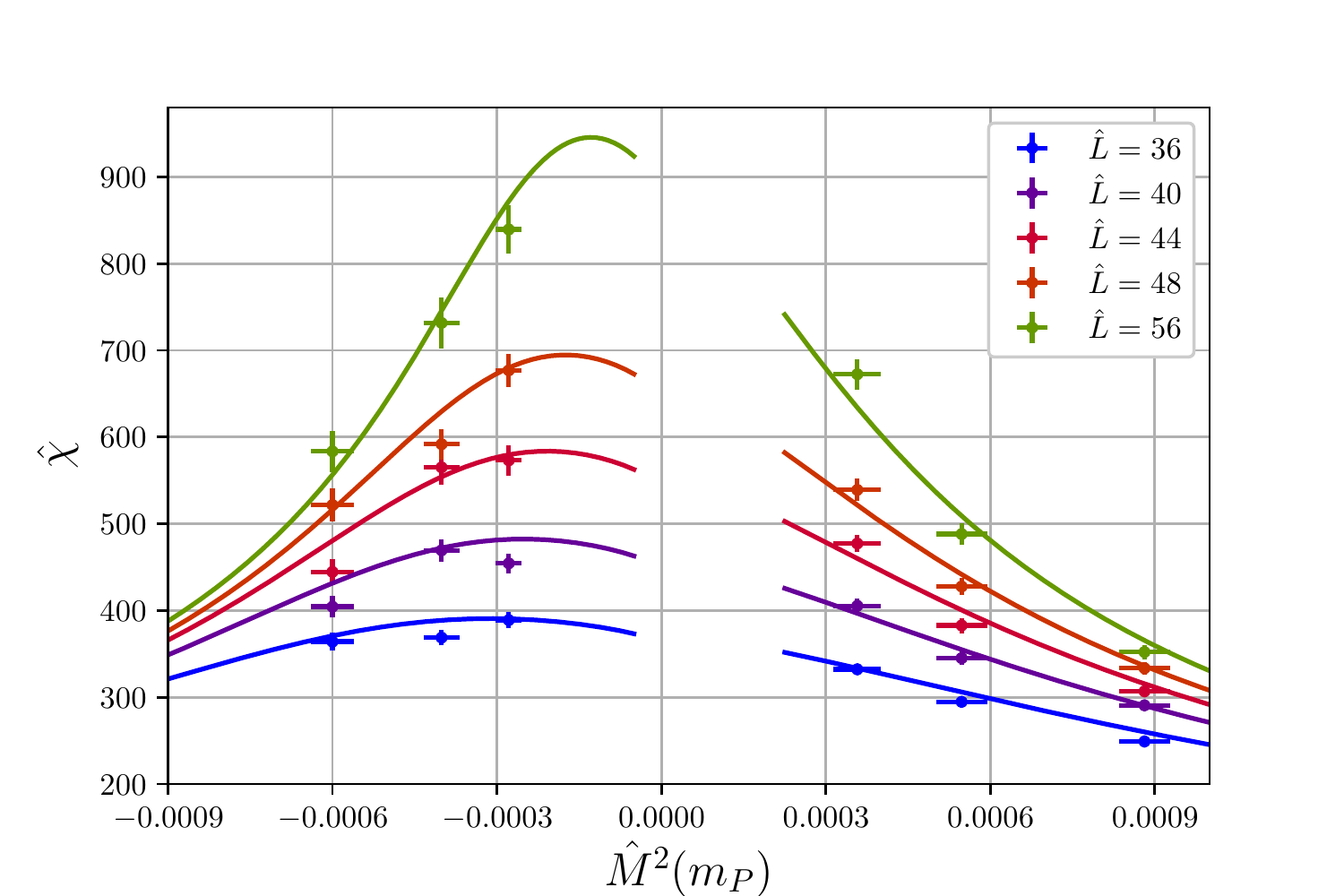}
\includegraphics[width=0.45\columnwidth]{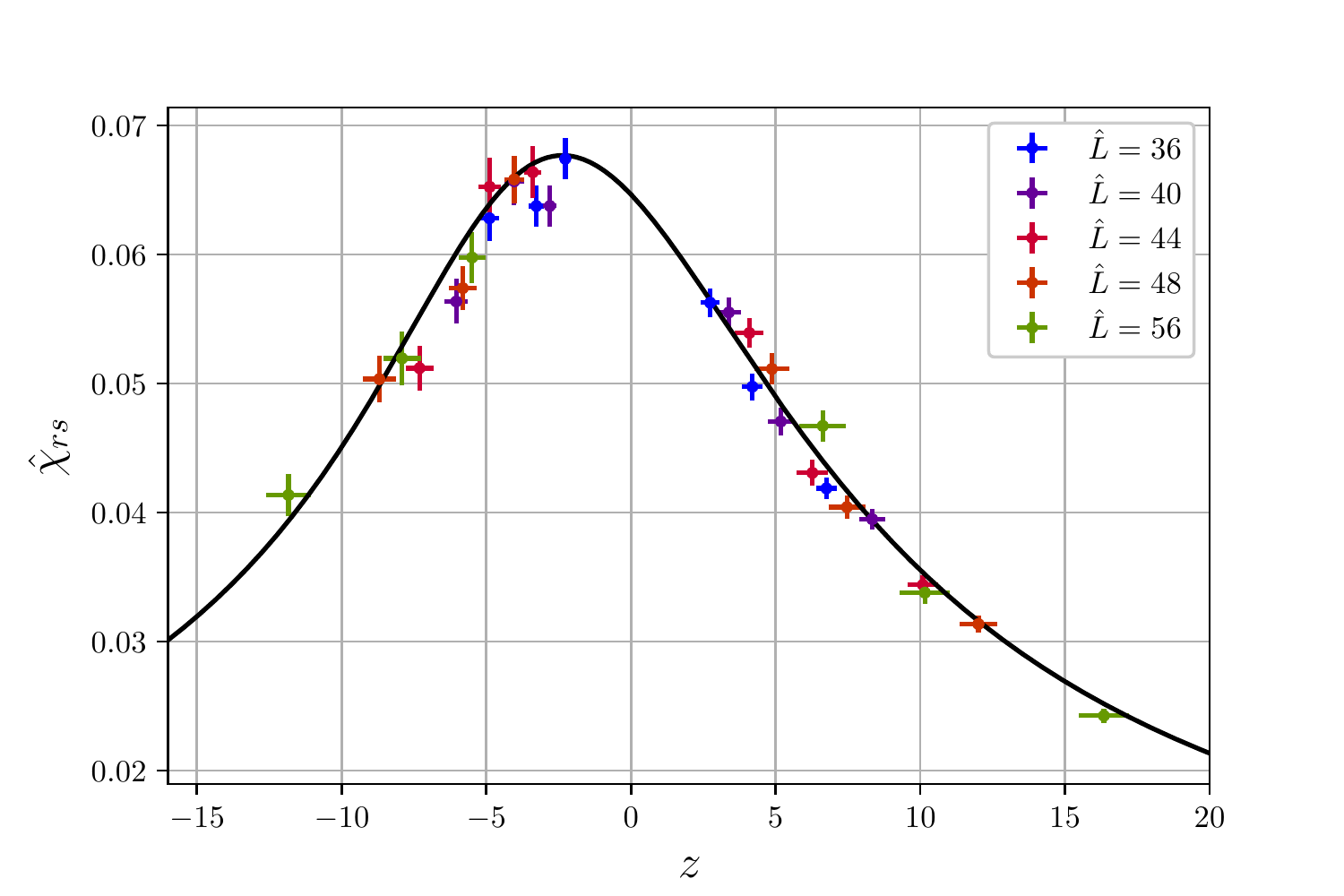}
\hspace*{0cm}(c)\hspace*{7.80cm}(d)\\
\includegraphics[width=0.45\columnwidth]{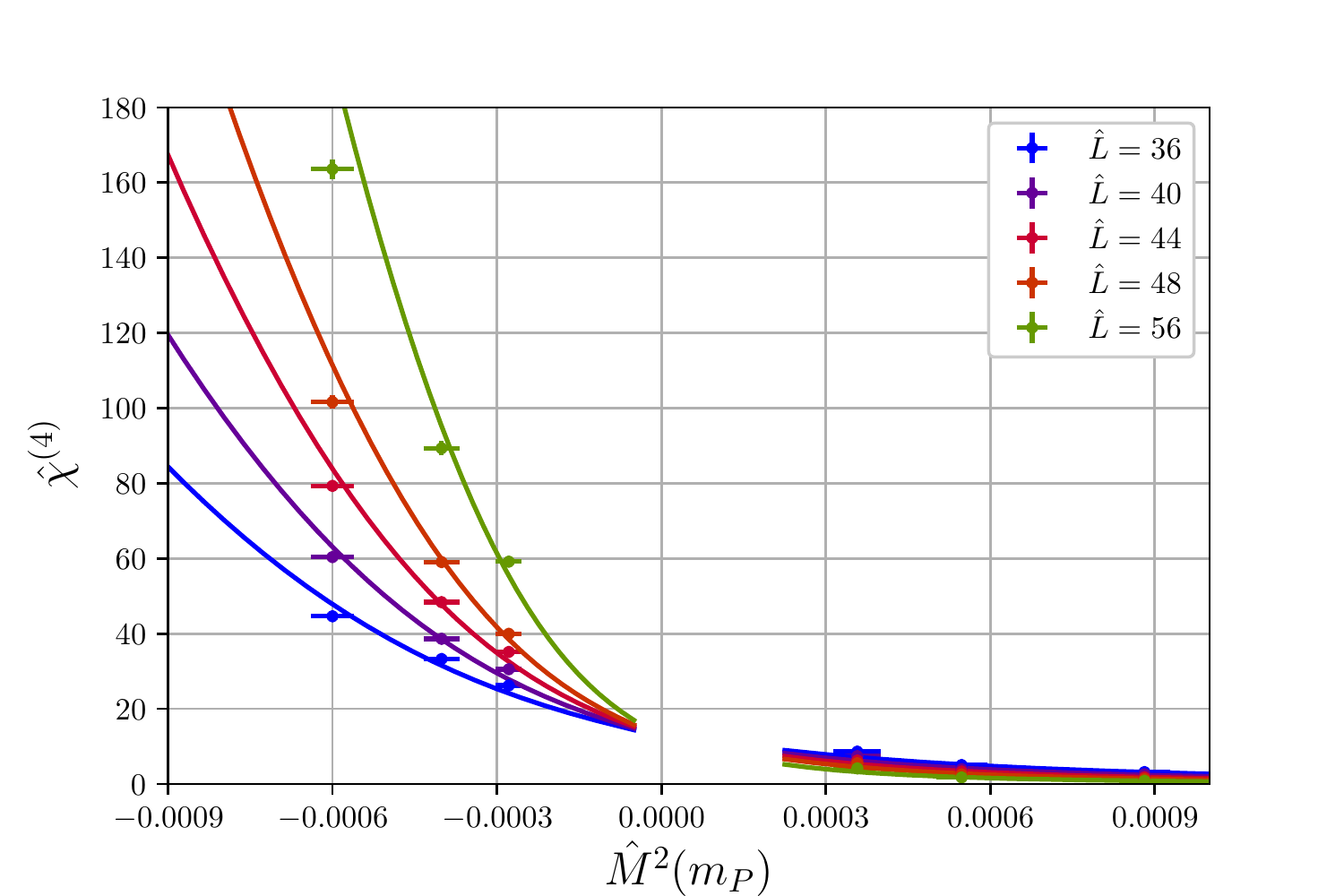}
\includegraphics[width=0.45\columnwidth]{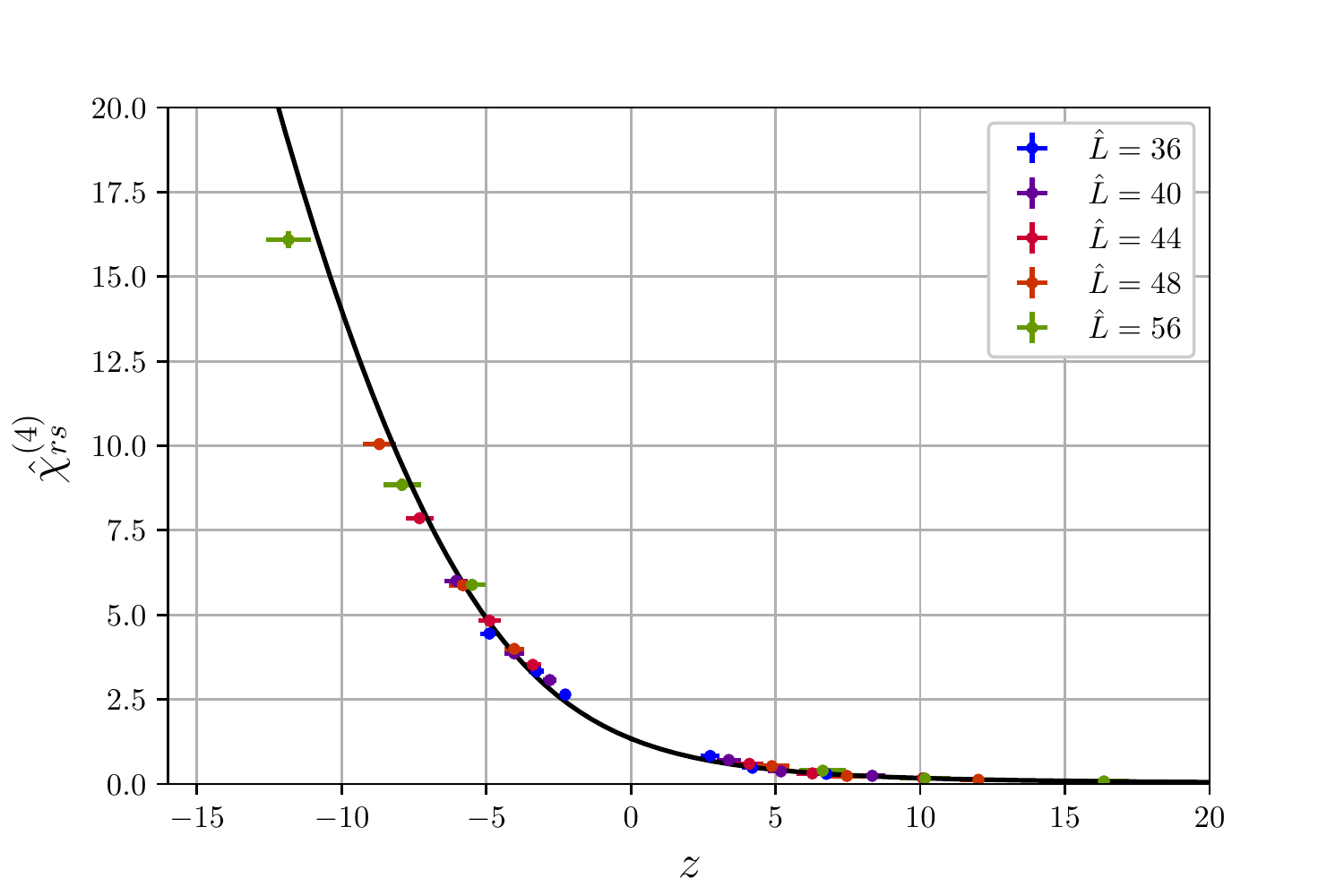}
\hspace*{0cm}(e)\hspace*{7.80cm}(f)
\caption[Global Scaling Test]{
Result of the scaling test {\it via } fitting $\la \hat{\varphi} \ra$, $\hat{\chi}$
and $\hat{\chi}^{(4)}$ simultaneously to Eqs.~(\ref{eq:scaling_mag}),
(\ref{eq:scaling_sus}) and (\ref{eq:scaling_fourPT}) are displayed in
(a), (c) and (e).  The corresponding rescaled quantities defined in Eqs.~(\ref{eq:rescaled_mag}),
(\ref{eq:rescaled_sus}) and (\ref{eq:rescaled_fourPT}) are plotted
against the scaling variable in (b), (d) and (f).
}
\label{fig:Global_fits}
\end{center}
\end{figure}

To have more stringent test for our scaling formulae, we also perform simultaneous
fits to $\la \hat{\varphi} \ra$, $\hat{\chi}$ and
$\hat{\chi}^{(4)}$.  Results of this analysis procedure are displayed
in Fig.~\ref{fig:Global_fits}.  
Compared to the individual fits discussed in the last paragraph, it is
more challenging to have good $\chi^{2}_{r}$ $(\equiv \chi^{2}/{\mathrm{dof}})$ in this
``global fit'', because of a much larger number of data points involved.
Nevertheless, we still find reasonable $\chi^{2}_{r}$ in
this procedure, and the scaling behaviour is also observed.
Furthermore the fit parameters, $A_{sy}$,
$\lambda_{sy}$, $A_{br}$ and $\lambda_{br}$, determined using this
method are comparable with those extracted from the individual-fit analysis.

Table~\ref{tab:summary_of_fits} lists the parameters, $A_{sy}$,
$\lambda_{sy}$, $A_{br}$ and $\lambda_{br}$, together with $\chi^{2}_{r}$, obtained from the above strategies.
\begin{table}[t]
\centering
    \begin{tabular}{ c | c | c | c | c | c }

                          & $\lambda_{sy}$ & $\lambda_{br}$ & $A_{sy}$ & $A_{br}$ & $\chi^{2}_{r}$ \\ \hline
        $\la \hat{\varphi} \ra$  & $0.0512 (52)$ & $0.0647 (60)$ & $0.873 (19)$ & $0.926 (24)$ & $1.37$ \\ 
        without Logs   & $0.0539 (57)$ & $0.0618 (54)$ & $0.878 (20)$ & $0.910 (22)$ & $1.35$ \\ \hline
        $\hat{\chi}$ & $0.0576 (64)$ & $0.0523 (57)$ & $0.875 (20)$ & $0.837 (21)$ & $0.91$ \\ 
        without Logs   & $0.0605 (71)$ & $0.0541 (62)$ & $0.882 (21)$ & $0.843 (22)$ & $0.90$\\ \hline
        $\hat{\chi}^{(4)}$ & $0.0512 (31)$ & $0.0651 (29)$ & $0.876 (5)$ & $0.930 (16)$ & $1.36$\\ 
        without Logs   & $0.0538 (34)$ & $0.0621 (26)$ & $0.880 (5)$ & $0.914 (15)$ & $1.34$ \\ \hline
        $\la \hat{\varphi} \ra$ and $\hat{\chi}^{(4)}$ & $0.0513 (30)$ & $0.0649 (29)$ & $0.875 (5)$ & $0.928 (16)$ & $1.32$ \\ 
        without Logs   & $0.0539 (34)$ & $0.0619 (26)$ & $0.880 (5)$ & $0.913 (15)$ & $1.30$ \\ \hline
        $\la \hat{\varphi} \ra$ and $\hat{\chi}$ & $0.0582 (22)$ & $0.0494 (9)$ & $0.880 (4)$ & $0.833 (6)$ & $2.08$ \\
        without Logs   & $0.0610 (24)$ & $0.0502 (9)$ &  $0.886 (9)$ & $0.837(6)$ & $1.84$ \\ \hline
        $\hat{\chi}$ and $\hat{\chi}^{(4)}$ & $0.0596 (22)$ & $0.0490 (9)$ & $0.884 (4)$ & $0.832 (6)$ & $2.24$ \\ 
    without Logs   & $0.0627 (25)$ & $0.0499 (9)$ & $0.890 (4)$ & $0.837 (6)$ & $1.98$ \\ \hline
        Global fit     & $0.0572 (23)$ & $0.0506 (9)$ & $0.880 (4)$ & $0.842 (6)$ & $2.29$\\ 
        without Logs   & $0.0601 (25)$ & $0.0514 (10)$ & $0.887 (4)$ & $0.847 (6)$ &  $2.01$ 
    
    \end{tabular} 
    \caption[Summary of all the scaling fit tests]{
    A summary of all fit parameters from all scaling tests performed: individual,  
global and combination of two observables. 
    \label{tab:summary_of_fits}
    }
\end{table}
Also tabulated are results from
simultaneous FSS fits for two observables amongst $\la \hat{\varphi} \ra$, $\hat{\chi}$
and $\hat{\chi}^{(4)}$.  This table shows that all these fits
find an overall agreement on the values of $A_{sy}$,
$\lambda_{sy}$, $A_{br}$ and $\lambda_{br}$.  In addition, the $\chi^2_r$
is always acceptable, suggesting that our scaling formulae describe 
the numerical data in a satisfactory fashion.   Finally, 
we have also tried the scaling tests
without logarithms {\it i.e.}, by setting $\lambda(\hat{L}) = \lambda(\hat{m}_P^{-1})$ 
in the scaling formulae, Eqs.~(\ref{eq:scaling_mag}),
(\ref{eq:scaling_sus}) and (\ref{eq:scaling_fourPT}).  Results from
this procedure are in the rows marked as ``without Logs'' in
Table~\ref{tab:summary_of_fits}.   These results seem to indicate that
our data are still not sensitive to the logarithms.  To detect
these logarithmic volume-dependence, even larger lattices and an
increased accuracy need to be reached.   
The fact that we are finding it very challenging
    to identify LL corrections to the mean-field scaling laws is consistent with results
    from a recent study of the $Z_{2}$ scalar field
    theory presented in Ref.~\cite{Hogervorst:2011zw}, where the
    authors adopt a slightly different approach from ours.
Here we stress that although our data do not
    allow us to discern the LL contributions in the scaling laws,
    the test presented in this section is already offering non-trivial
    information.  As discussed in Sec.~\ref{sec:scaling_formulae}, 
     the use of Eqs.~(\ref{eq:varphi_RG_running}) and
    (\ref{recursion_formula}) in obtaining the FSS formulae is
    justified only for the Gaussian fixed point.  This aspect of the
    derivation is identical for the Higgs-Yukawa and the pure-scalar
    O(4) models.  In other words, the scaling behaviour of $\la
    \hat{\varphi} \ra$, $\hat{\chi}$ and $\hat{\chi}_{4}$ in these two
    models take the same functional form in Eqs.~(\ref{eq:scaling_mag}), (\ref{eq:scaling_sus})
    and (\ref{eq:scaling_fourPT}), with the difference being the
    explicit logarithmic volume-dependence encoded in $\lambda$ and
    $z$.  In view of this, the fact that our formulae fit the lattice
    data well is already indicating the validity of employing Eqs.~(\ref{eq:varphi_RG_running}) and
    (\ref{recursion_formula}) for the investigation of scaling
    behaviour governed by the Gaussian fixed point.  Furthermore, the pure-scalar O(4) theory can be regarded as the
Higgs-Yukawa model in the limit where the Yukawa coupling vanishes.  
As the first step, this is a meaningful limit to examine our formulae, since the next-stage
test of our analytic results should be performed in the Higgs-Yukawa
model with small Yukawa coupling.  
Therefore, this scalar-theory numerical exercise can be taken as a good prototype study
for our future investigation of the Higgs-Yukawa model.  Since
simulations for the Higgs-Yukawa model is very demanding in computing resources, it is especially
essential to perform such a prototype study to identify possible range of
parameters ({\it e.g.}, the lattice volumes needed) that would be needed to
observe scaling behaviour.

As discussed earlier in this subsection, our data indicate that within
our numerical accuracy, the field variables can be assumed to be already naturally
matched to a MOM scheme, leading to the introduction of $A(m^{-1}_{P})$ in
the scaling formulae to accommodate the matching from this MOM scheme
to the on-shell scheme.  To examine this assumption, we carry out the
fits by allowing $\kappa{-}$dependence (hence lattice-spacing-dependence) in $A(m^{-1}_{P})$.  These
$\kappa{-}$dependent $A(m^{-1}_{P})$ can be interpreted as the
matching coefficients between the bare lattice $\varphi$ and its
renormalised counterpart in the on-shell scheme.  It is found that
this procedure results in values of $A(m^{-1}_{P})$ and $\lambda
(m_{P}^{-1})$ that are well consistent with those listed in Table~\ref{tab:summary_of_fits}.

We conclude this section by stating the remark that we find the 
renormalised quartic coupling always to be about a factor 
of three smaller than the bare one. This shows a strong renormalisation 
effect in the expected direction, providing further evidence that our
strategy is valid.

\section{Conclusion}
\label{sec:conclusion}
In this work, we derive for the first time FSS formulae for the moments of the scalar-field zero mode in a four-dimensional Higgs-Yukawa model.  It is found that 
near the Gaussian fixed point, the scaling laws can be derived for these moments by following the strategy outlined in Ref.~\cite{Brezin:1985xx}.  
We discuss in detail the incorporation of the leading-logarithmic corrections to these scaling laws {\it via} solving the one-loop RG equations for the theory.   The solution to these equations involves integration constants which appear in the expression of the scalar zero-mode moments, and in the scaling variable.   These integration constants can be treated as free parameters that can in principle be extracted by fitting lattice-simulation data with our scaling formulae.  Through investigating the quality of such fits, one can determine whether or not particular critical points are associated with the Gaussian fixed point.  That is, our formulae can be used to establish the triviality property of the Higgs-Yukawa model at numerical accuracy that can be achieved, or to search for alternative scenarios in the phase structure of the theory.

To examine the scaling laws derived in this project,  we confront them with data from numerical simulations in lattice field theory.  Since the FSS behaviour of the Higgs-Yukawa model is rather complex, and large-volume simulations of the model are presently very expensive, in the current project we proceed to implement the test in a simpler model, namely the O(4) pure scalar theory.  In particular, we study the volume-dependence in three quantities: the magnetisation, the susceptibility and the fourth moment constructed from the zero mode of the scalar field.   Compared with the Higgs-Yukawa model, although the functional form of the FSS formulae described by Eqs.~(\ref{eq:varphi_RG_running}) and (\ref{recursion_formula}) remains the same in this pure scalar theory, the scaling variable simplifies significantly.  In addition to performing fits to the above observables separately, we investigate the global fit by determining the free parameters in the scaling behaviour from these three quantities simultaneously.  Reasonable fits and compatible results are obtained from both approaches, employing simulations carried out with a weak bare quartic coupling near the critical point at five values of the lattice size in the range $36 \le L/a \le 56$.  This shows that our scaling formulae are indeed valid.   On the other hand, we also discover that our data are not sensitive to the effects of the LL corrections to the mean-field scaling laws.  However, it is always challenging to concretely discern logarithmic behaviour in scaling tests, and we conclude that numerical calculations at bigger lattice volumes are needed for this aspect of the test, which is beyond the scope of the current work.

The above numerical analysis demonstrates that large lattice sizes can be essential for the scaling test of the Higgs-Yukawa model.  This means that future dedicated efforts and long-term research programmes will be necessary for this task.   It is, withal, crucial to have such a test for our understanding of non-perturbative aspects of the SM.  Our FSS formulae presented in Sec.~\ref{sec:scaling_formulae} of this article can eventually be employed to examine the fixed-point structure at large values of the Yukawa coupling~\cite{Hasenfratz:1992xs,Bulava:2012rb}. In addition, these formulae could be generalised to more flavours of quarks and possibly to different dimensions as well as the inclusion of gauge fields, making them applicable for similar studies of a broader class of Higgs-Yukawa systems in high-energy and condensed-matter physics~\cite{Shaposhnikov:2009pv,Gies:2013pma,Litim:2014uca,Maas:2017wzi,Eichhorn:2017eht}.


\acknowledgements
The authors thank Philipp Gerhold, Chung-Wen Kao and Attila Nagy for
useful discussions.   DYJC and CJDL are supported by Taiwanese MoST
{\it via} grant 105-2112-M-002-023-MY3.
Numerical calculations have
been carried out on the HPC facilities at National Chiao-Tung
University and DESY Zeuthen computer centre, as well as
on SGI system HLRN-II at HLRN supercomputing service Berlin-Hannover.



\clearpage

\bibliographystyle{apsrev.bst} 
\bibliography{refs} 

\begin{thebibliography}{81}
\expandafter\ifx\csname natexlab\endcsname\relax\def\natexlab#1{#1}\fi
\expandafter\ifx\csname bibnamefont\endcsname\relax
  \def\bibnamefont#1{#1}\fi
\expandafter\ifx\csname bibfnamefont\endcsname\relax
  \def\bibfnamefont#1{#1}\fi
\expandafter\ifx\csname citenamefont\endcsname\relax
  \def\citenamefont#1{#1}\fi
\expandafter\ifx\csname url\endcsname\relax
  \def\url#1{\texttt{#1}}\fi
\expandafter\ifx\csname urlprefix\endcsname\relax\def\urlprefix{URL }\fi
\providecommand{\bibinfo}[2]{#2}
\providecommand{\eprint}[2][]{\url{#2}}

\bibitem[{\citenamefont{Aizenman}(1981)}]{Aizenman:1981zz}
\bibinfo{author}{\bibfnamefont{M.}~\bibnamefont{Aizenman}},
  \bibinfo{journal}{Phys.Rev.Lett.} \textbf{\bibinfo{volume}{47}},
  \bibinfo{pages}{886} (\bibinfo{year}{1981}).

\bibitem[{\citenamefont{Frohlich}(1982)}]{Frohlich:1982tw}
\bibinfo{author}{\bibfnamefont{J.}~\bibnamefont{Frohlich}},
  \bibinfo{journal}{Nucl.Phys.} \textbf{\bibinfo{volume}{B200}},
  \bibinfo{pages}{281} (\bibinfo{year}{1982}).

\bibitem[{\citenamefont{Luscher and Weisz}(1987)}]{Luscher:1987ay}
\bibinfo{author}{\bibfnamefont{M.}~\bibnamefont{Luscher}} \bibnamefont{and}
  \bibinfo{author}{\bibfnamefont{P.}~\bibnamefont{Weisz}},
  \bibinfo{journal}{Nucl. Phys.} \textbf{\bibinfo{volume}{B290}},
  \bibinfo{pages}{25} (\bibinfo{year}{1987}).

\bibitem[{\citenamefont{Luscher and
  Weisz}(1988{\natexlab{a}})}]{Luscher:1987ek}
\bibinfo{author}{\bibfnamefont{M.}~\bibnamefont{Luscher}} \bibnamefont{and}
  \bibinfo{author}{\bibfnamefont{P.}~\bibnamefont{Weisz}},
  \bibinfo{journal}{Nucl. Phys.} \textbf{\bibinfo{volume}{B295}},
  \bibinfo{pages}{65} (\bibinfo{year}{1988}{\natexlab{a}}).

\bibitem[{\citenamefont{Luscher and Weisz}(1989)}]{Luscher:1988uq}
\bibinfo{author}{\bibfnamefont{M.}~\bibnamefont{Luscher}} \bibnamefont{and}
  \bibinfo{author}{\bibfnamefont{P.}~\bibnamefont{Weisz}},
  \bibinfo{journal}{Nucl. Phys.} \textbf{\bibinfo{volume}{B318}},
  \bibinfo{pages}{705} (\bibinfo{year}{1989}).

\bibitem[{\citenamefont{Bernreuther and Gockeler}(1988)}]{Bernreuther:1987hv}
\bibinfo{author}{\bibfnamefont{W.}~\bibnamefont{Bernreuther}} \bibnamefont{and}
  \bibinfo{author}{\bibfnamefont{M.}~\bibnamefont{Gockeler}},
  \bibinfo{journal}{Nucl.Phys.} \textbf{\bibinfo{volume}{B295}},
  \bibinfo{pages}{199} (\bibinfo{year}{1988}).

\bibitem[{\citenamefont{Kenna and Lang}(1993)}]{Kenna:1992np}
\bibinfo{author}{\bibfnamefont{R.}~\bibnamefont{Kenna}} \bibnamefont{and}
  \bibinfo{author}{\bibfnamefont{C.}~\bibnamefont{Lang}},
  \bibinfo{journal}{Nucl.Phys.} \textbf{\bibinfo{volume}{B393}},
  \bibinfo{pages}{461} (\bibinfo{year}{1993}).

\bibitem[{\citenamefont{Gockeler et~al.}(1993)\citenamefont{Gockeler, Kastrup,
  Neuhaus, and Zimmermann}}]{Gockeler:1992zj}
\bibinfo{author}{\bibfnamefont{M.}~\bibnamefont{Gockeler}},
  \bibinfo{author}{\bibfnamefont{H.~A.} \bibnamefont{Kastrup}},
  \bibinfo{author}{\bibfnamefont{T.}~\bibnamefont{Neuhaus}}, \bibnamefont{and}
  \bibinfo{author}{\bibfnamefont{F.}~\bibnamefont{Zimmermann}},
  \bibinfo{journal}{Nucl. Phys.} \textbf{\bibinfo{volume}{B404}},
  \bibinfo{pages}{517} (\bibinfo{year}{1993}), \eprint{hep-lat/9206025}.

\bibitem[{\citenamefont{Wolff}(2009)}]{Wolff:2009ke}
\bibinfo{author}{\bibfnamefont{U.}~\bibnamefont{Wolff}},
  \bibinfo{journal}{Phys. Rev.} \textbf{\bibinfo{volume}{D79}},
  \bibinfo{pages}{105002} (\bibinfo{year}{2009}), \eprint{0902.3100}.

\bibitem[{\citenamefont{Weisz and Wolff}(2011)}]{Weisz:2010xx}
\bibinfo{author}{\bibfnamefont{P.}~\bibnamefont{Weisz}} \bibnamefont{and}
  \bibinfo{author}{\bibfnamefont{U.}~\bibnamefont{Wolff}},
  \bibinfo{journal}{Nucl. Phys.} \textbf{\bibinfo{volume}{B846}},
  \bibinfo{pages}{316} (\bibinfo{year}{2011}), \eprint{1012.0404}.

\bibitem[{\citenamefont{Hogervorst and Wolff}(2012)}]{Hogervorst:2011zw}
\bibinfo{author}{\bibfnamefont{M.}~\bibnamefont{Hogervorst}} \bibnamefont{and}
  \bibinfo{author}{\bibfnamefont{U.}~\bibnamefont{Wolff}},
  \bibinfo{journal}{Nucl. Phys.} \textbf{\bibinfo{volume}{B855}},
  \bibinfo{pages}{885} (\bibinfo{year}{2012}), \eprint{1109.6186}.

\bibitem[{\citenamefont{Siefert and Wolff}(2014)}]{Siefert:2014ela}
\bibinfo{author}{\bibfnamefont{J.}~\bibnamefont{Siefert}} \bibnamefont{and}
  \bibinfo{author}{\bibfnamefont{U.}~\bibnamefont{Wolff}},
  \bibinfo{journal}{Phys. Lett.} \textbf{\bibinfo{volume}{B733}},
  \bibinfo{pages}{11} (\bibinfo{year}{2014}), \eprint{1403.2570}.

\bibitem[{\citenamefont{Dashen and Neuberger}(1983)}]{Dashen:1983ts}
\bibinfo{author}{\bibfnamefont{R.~F.} \bibnamefont{Dashen}} \bibnamefont{and}
  \bibinfo{author}{\bibfnamefont{H.}~\bibnamefont{Neuberger}},
  \bibinfo{journal}{Phys.Rev.Lett.} \textbf{\bibinfo{volume}{50}},
  \bibinfo{pages}{1897} (\bibinfo{year}{1983}).

\bibitem[{\citenamefont{Hasenfratz et~al.}(1987)\citenamefont{Hasenfratz,
  Jansen, Lang, Neuhaus, and Yoneyama}}]{Hasenfratz:1987eh}
\bibinfo{author}{\bibfnamefont{A.}~\bibnamefont{Hasenfratz}},
  \bibinfo{author}{\bibfnamefont{K.}~\bibnamefont{Jansen}},
  \bibinfo{author}{\bibfnamefont{C.~B.} \bibnamefont{Lang}},
  \bibinfo{author}{\bibfnamefont{T.}~\bibnamefont{Neuhaus}}, \bibnamefont{and}
  \bibinfo{author}{\bibfnamefont{H.}~\bibnamefont{Yoneyama}},
  \bibinfo{journal}{Phys.Lett.} \textbf{\bibinfo{volume}{B199}},
  \bibinfo{pages}{531} (\bibinfo{year}{1987}).

\bibitem[{\citenamefont{Kuti et~al.}(1988)\citenamefont{Kuti, Lin, and
  Shen}}]{Kuti:1987nr}
\bibinfo{author}{\bibfnamefont{J.}~\bibnamefont{Kuti}},
  \bibinfo{author}{\bibfnamefont{L.}~\bibnamefont{Lin}}, \bibnamefont{and}
  \bibinfo{author}{\bibfnamefont{Y.}~\bibnamefont{Shen}},
  \bibinfo{journal}{Phys.Rev.Lett.} \textbf{\bibinfo{volume}{61}},
  \bibinfo{pages}{678} (\bibinfo{year}{1988}).

\bibitem[{\citenamefont{Luscher and
  Weisz}(1988{\natexlab{b}})}]{Luscher:1988gc}
\bibinfo{author}{\bibfnamefont{M.}~\bibnamefont{Luscher}} \bibnamefont{and}
  \bibinfo{author}{\bibfnamefont{P.}~\bibnamefont{Weisz}},
  \bibinfo{journal}{Phys.Lett.} \textbf{\bibinfo{volume}{B212}},
  \bibinfo{pages}{472} (\bibinfo{year}{1988}{\natexlab{b}}).

\bibitem[{\citenamefont{Branchina and Messina}(2013)}]{Branchina:2013jra}
\bibinfo{author}{\bibfnamefont{V.}~\bibnamefont{Branchina}} \bibnamefont{and}
  \bibinfo{author}{\bibfnamefont{E.}~\bibnamefont{Messina}},
  \bibinfo{journal}{Phys.Rev.Lett.} \textbf{\bibinfo{volume}{111}},
  \bibinfo{pages}{241801} (\bibinfo{year}{2013}).

\bibitem[{\citenamefont{Gies et~al.}(2014)\citenamefont{Gies, Gneiting, and
  Sondenheimer}}]{Gies:2013fua}
\bibinfo{author}{\bibfnamefont{H.}~\bibnamefont{Gies}},
  \bibinfo{author}{\bibfnamefont{C.}~\bibnamefont{Gneiting}}, \bibnamefont{and}
  \bibinfo{author}{\bibfnamefont{R.}~\bibnamefont{Sondenheimer}},
  \bibinfo{journal}{Phys.Rev.} \textbf{\bibinfo{volume}{D89}},
  \bibinfo{pages}{045012} (\bibinfo{year}{2014}).

\bibitem[{\citenamefont{Chu et~al.}(2015)\citenamefont{Chu, Jansen,
  Knippschild, Lin, and Nagy}}]{Chu:2015nha}
\bibinfo{author}{\bibfnamefont{D.~Y.-J.} \bibnamefont{Chu}},
  \bibinfo{author}{\bibfnamefont{K.}~\bibnamefont{Jansen}},
  \bibinfo{author}{\bibfnamefont{B.}~\bibnamefont{Knippschild}},
  \bibinfo{author}{\bibfnamefont{C.-J.~D.} \bibnamefont{Lin}},
  \bibnamefont{and} \bibinfo{author}{\bibfnamefont{A.}~\bibnamefont{Nagy}},
  \bibinfo{journal}{Phys.Lett.} \textbf{\bibinfo{volume}{B744}},
  \bibinfo{pages}{146} (\bibinfo{year}{2015}).

\bibitem[{\citenamefont{{J. Brehmer, A. Freitas, D. L\'{o}pez-Val and T.
  Plehn}}(2016)}]{Brehmer:2015rna}
\bibinfo{author}{\bibnamefont{{J. Brehmer, A. Freitas, D. L\'{o}pez-Val and T.
  Plehn}}}, \bibinfo{journal}{Phys. Rev.} \textbf{\bibinfo{volume}{D93}},
  \bibinfo{pages}{075014} (\bibinfo{year}{2016}), \eprint{1510.03443}.

\bibitem[{\citenamefont{{A. Biek\"{o}tter, J. Brehmer, Johann and T.
  Plehn}}(2016)}]{Biekotter:2016ecg}
\bibinfo{author}{\bibnamefont{{A. Biek\"{o}tter, J. Brehmer, Johann and T.
  Plehn}}}, \bibinfo{journal}{Phys. Rev.} \textbf{\bibinfo{volume}{D94}},
  \bibinfo{pages}{055032} (\bibinfo{year}{2016}), \eprint{1602.05202}.

\bibitem[{\citenamefont{Chu et~al.}(2018)\citenamefont{Chu, Jansen,
  Knippschild, and Lin}}]{Chu:2017vmc}
\bibinfo{author}{\bibfnamefont{D.~Y.~J.} \bibnamefont{Chu}},
  \bibinfo{author}{\bibfnamefont{K.}~\bibnamefont{Jansen}},
  \bibinfo{author}{\bibfnamefont{B.}~\bibnamefont{Knippschild}},
  \bibnamefont{and} \bibinfo{author}{\bibfnamefont{C.~J.~D.}
  \bibnamefont{Lin}}, \bibinfo{journal}{EPJ Web Conf.}
  \textbf{\bibinfo{volume}{175}}, \bibinfo{pages}{08017}
  (\bibinfo{year}{2018}), \eprint{1710.09737}.

\bibitem[{\citenamefont{Dawson}(2017)}]{Dawson:2017ksx}
\bibinfo{author}{\bibfnamefont{S.}~\bibnamefont{Dawson}}, in
  \emph{\bibinfo{booktitle}{{Proceedings, Theoretical Advanced Study Institute
  in Elementary Particle Physics : Anticipating the Next Discoveries in
  Particle Physics (TASI 2016): Boulder, CO, USA, June 6-July 1, 2016}}}
  (\bibinfo{year}{2017}), pp. \bibinfo{pages}{1--63}, \eprint{1712.07232}.

\bibitem[{\citenamefont{Holdom}(1985)}]{Holdom:1984sk}
\bibinfo{author}{\bibfnamefont{B.}~\bibnamefont{Holdom}},
  \bibinfo{journal}{Phys. Lett.} \textbf{\bibinfo{volume}{150B}},
  \bibinfo{pages}{301} (\bibinfo{year}{1985}).

\bibitem[{\citenamefont{Yamawaki et~al.}(1986)\citenamefont{Yamawaki, Bando,
  and Matumoto}}]{Yamawaki:1985zg}
\bibinfo{author}{\bibfnamefont{K.}~\bibnamefont{Yamawaki}},
  \bibinfo{author}{\bibfnamefont{M.}~\bibnamefont{Bando}}, \bibnamefont{and}
  \bibinfo{author}{\bibfnamefont{K.-i.} \bibnamefont{Matumoto}},
  \bibinfo{journal}{Phys. Rev. Lett.} \textbf{\bibinfo{volume}{56}},
  \bibinfo{pages}{1335} (\bibinfo{year}{1986}).

\bibitem[{\citenamefont{Appelquist et~al.}(1986)\citenamefont{Appelquist,
  Karabali, and Wijewardhana}}]{Appelquist:1986an}
\bibinfo{author}{\bibfnamefont{T.~W.} \bibnamefont{Appelquist}},
  \bibinfo{author}{\bibfnamefont{D.}~\bibnamefont{Karabali}}, \bibnamefont{and}
  \bibinfo{author}{\bibfnamefont{L.~C.~R.} \bibnamefont{Wijewardhana}},
  \bibinfo{journal}{Phys. Rev. Lett.} \textbf{\bibinfo{volume}{57}},
  \bibinfo{pages}{957} (\bibinfo{year}{1986}).

\bibitem[{\citenamefont{Lucini}()}]{Lucini:2015noa}
\bibinfo{author}{\bibfnamefont{B.}~\bibnamefont{Lucini}},
  \bibinfo{note}{arXiv:1503.00371}.

\bibitem[{\citenamefont{Nogradi and Patella}(2016)}]{Nogradi:2016qek}
\bibinfo{author}{\bibfnamefont{D.}~\bibnamefont{Nogradi}} \bibnamefont{and}
  \bibinfo{author}{\bibfnamefont{A.}~\bibnamefont{Patella}},
  \bibinfo{journal}{Int. J. Mod. Phys.} \textbf{\bibinfo{volume}{A31}},
  \bibinfo{pages}{1643003} (\bibinfo{year}{2016}), \eprint{1607.07638}.

\bibitem[{\citenamefont{Svetitsky}(2018)}]{Svetitsky:2017xqk}
\bibinfo{author}{\bibfnamefont{B.}~\bibnamefont{Svetitsky}},
  \bibinfo{journal}{EPJ Web Conf.} \textbf{\bibinfo{volume}{175}},
  \bibinfo{pages}{01017} (\bibinfo{year}{2018}), \eprint{1708.04840}.

\bibitem[{\citenamefont{Piai}(2010)}]{Piai:2010ma}
\bibinfo{author}{\bibfnamefont{M.}~\bibnamefont{Piai}},
  \bibinfo{journal}{Adv.High Energy Phys.} \textbf{\bibinfo{volume}{2010}},
  \bibinfo{pages}{464302} (\bibinfo{year}{2010}).

\bibitem[{\citenamefont{Piai}(2014)}]{Piai:2014pla}
\bibinfo{author}{\bibfnamefont{M.}~\bibnamefont{Piai}}, in
  \emph{\bibinfo{booktitle}{{Proceedings, KMI-GCOE Workshop on Strong Coupling
  Gauge Theories in the LHC Perspective (SCGT 12): Nagoya, Japan, December 4-7,
  2012}}} (\bibinfo{year}{2014}), pp. \bibinfo{pages}{185--191}.

\bibitem[{\citenamefont{Kaplan et~al.}(1984)\citenamefont{Kaplan, Georgi, and
  Dimopoulos}}]{Kaplan:1983sm}
\bibinfo{author}{\bibfnamefont{D.~B.} \bibnamefont{Kaplan}},
  \bibinfo{author}{\bibfnamefont{H.}~\bibnamefont{Georgi}}, \bibnamefont{and}
  \bibinfo{author}{\bibfnamefont{S.}~\bibnamefont{Dimopoulos}},
  \bibinfo{journal}{Phys. Lett.} \textbf{\bibinfo{volume}{136B}},
  \bibinfo{pages}{187} (\bibinfo{year}{1984}).

\bibitem[{\citenamefont{Georgi and Kaplan}(1984)}]{Georgi:1984af}
\bibinfo{author}{\bibfnamefont{H.}~\bibnamefont{Georgi}} \bibnamefont{and}
  \bibinfo{author}{\bibfnamefont{D.~B.} \bibnamefont{Kaplan}},
  \bibinfo{journal}{Phys. Lett.} \textbf{\bibinfo{volume}{145B}},
  \bibinfo{pages}{216} (\bibinfo{year}{1984}).

\bibitem[{\citenamefont{Dugan et~al.}(1985)\citenamefont{Dugan, Georgi, and
  Kaplan}}]{Dugan:1984hq}
\bibinfo{author}{\bibfnamefont{M.~J.} \bibnamefont{Dugan}},
  \bibinfo{author}{\bibfnamefont{H.}~\bibnamefont{Georgi}}, \bibnamefont{and}
  \bibinfo{author}{\bibfnamefont{D.~B.} \bibnamefont{Kaplan}},
  \bibinfo{journal}{Nucl. Phys.} \textbf{\bibinfo{volume}{B254}},
  \bibinfo{pages}{299} (\bibinfo{year}{1985}).

\bibitem[{\citenamefont{Barnard et~al.}(2014)\citenamefont{Barnard, Gherghetta,
  and Ray}}]{Barnard:2013zea}
\bibinfo{author}{\bibfnamefont{J.}~\bibnamefont{Barnard}},
  \bibinfo{author}{\bibfnamefont{T.}~\bibnamefont{Gherghetta}},
  \bibnamefont{and} \bibinfo{author}{\bibfnamefont{T.~S.} \bibnamefont{Ray}},
  \bibinfo{journal}{JHEP} \textbf{\bibinfo{volume}{02}}, \bibinfo{pages}{002}
  (\bibinfo{year}{2014}), \eprint{1311.6562}.

\bibitem[{\citenamefont{Ferretti and Karateev}(2014)}]{Ferretti:2013kya}
\bibinfo{author}{\bibfnamefont{G.}~\bibnamefont{Ferretti}} \bibnamefont{and}
  \bibinfo{author}{\bibfnamefont{D.}~\bibnamefont{Karateev}},
  \bibinfo{journal}{JHEP} \textbf{\bibinfo{volume}{03}}, \bibinfo{pages}{077}
  (\bibinfo{year}{2014}), \eprint{1312.5330}.

\bibitem[{\citenamefont{Cacciapaglia and Sannino}(2014)}]{Cacciapaglia:2014uja}
\bibinfo{author}{\bibfnamefont{G.}~\bibnamefont{Cacciapaglia}}
  \bibnamefont{and} \bibinfo{author}{\bibfnamefont{F.}~\bibnamefont{Sannino}},
  \bibinfo{journal}{JHEP} \textbf{\bibinfo{volume}{04}}, \bibinfo{pages}{111}
  (\bibinfo{year}{2014}), \eprint{1402.0233}.

\bibitem[{\citenamefont{Belyaev et~al.}(2017)\citenamefont{Belyaev,
  Cacciapaglia, Cai, Ferretti, Flacke, Parolini, and
  Serodio}}]{Belyaev:2016ftv}
\bibinfo{author}{\bibfnamefont{A.}~\bibnamefont{Belyaev}},
  \bibinfo{author}{\bibfnamefont{G.}~\bibnamefont{Cacciapaglia}},
  \bibinfo{author}{\bibfnamefont{H.}~\bibnamefont{Cai}},
  \bibinfo{author}{\bibfnamefont{G.}~\bibnamefont{Ferretti}},
  \bibinfo{author}{\bibfnamefont{T.}~\bibnamefont{Flacke}},
  \bibinfo{author}{\bibfnamefont{A.}~\bibnamefont{Parolini}}, \bibnamefont{and}
  \bibinfo{author}{\bibfnamefont{H.}~\bibnamefont{Serodio}},
  \bibinfo{journal}{JHEP} \textbf{\bibinfo{volume}{01}}, \bibinfo{pages}{094}
  (\bibinfo{year}{2017}), \bibinfo{note}{[Erratum: JHEP12,088(2017)]},
  \eprint{1610.06591}.

\bibitem[{\citenamefont{DeGrand et~al.}(2015)\citenamefont{DeGrand, Liu, Neil,
  Shamir, and Svetitsky}}]{DeGrand:2015lna}
\bibinfo{author}{\bibfnamefont{T.}~\bibnamefont{DeGrand}},
  \bibinfo{author}{\bibfnamefont{Y.}~\bibnamefont{Liu}},
  \bibinfo{author}{\bibfnamefont{E.~T.} \bibnamefont{Neil}},
  \bibinfo{author}{\bibfnamefont{Y.}~\bibnamefont{Shamir}}, \bibnamefont{and}
  \bibinfo{author}{\bibfnamefont{B.}~\bibnamefont{Svetitsky}},
  \bibinfo{journal}{Phys. Rev.} \textbf{\bibinfo{volume}{D91}},
  \bibinfo{pages}{114502} (\bibinfo{year}{2015}), \eprint{1501.05665}.

\bibitem[{\citenamefont{Ayyar et~al.}(2018{\natexlab{a}})\citenamefont{Ayyar,
  DeGrand, Golterman, Hackett, Jay, Neil, Shamir, and
  Svetitsky}}]{Ayyar:2017qdf}
\bibinfo{author}{\bibfnamefont{V.}~\bibnamefont{Ayyar}},
  \bibinfo{author}{\bibfnamefont{T.}~\bibnamefont{DeGrand}},
  \bibinfo{author}{\bibfnamefont{M.}~\bibnamefont{Golterman}},
  \bibinfo{author}{\bibfnamefont{D.~C.} \bibnamefont{Hackett}},
  \bibinfo{author}{\bibfnamefont{W.~I.} \bibnamefont{Jay}},
  \bibinfo{author}{\bibfnamefont{E.~T.} \bibnamefont{Neil}},
  \bibinfo{author}{\bibfnamefont{Y.}~\bibnamefont{Shamir}}, \bibnamefont{and}
  \bibinfo{author}{\bibfnamefont{B.}~\bibnamefont{Svetitsky}},
  \bibinfo{journal}{Phys. Rev.} \textbf{\bibinfo{volume}{D97}},
  \bibinfo{pages}{074505} (\bibinfo{year}{2018}{\natexlab{a}}),
  \eprint{1710.00806}.

\bibitem[{\citenamefont{Ayyar et~al.}(2018{\natexlab{b}})\citenamefont{Ayyar,
  Degrand, Hackett, Jay, Neil, Shamir, and Svetitsky}}]{Ayyar:2018zuk}
\bibinfo{author}{\bibfnamefont{V.}~\bibnamefont{Ayyar}},
  \bibinfo{author}{\bibfnamefont{T.}~\bibnamefont{Degrand}},
  \bibinfo{author}{\bibfnamefont{D.~C.} \bibnamefont{Hackett}},
  \bibinfo{author}{\bibfnamefont{W.~I.} \bibnamefont{Jay}},
  \bibinfo{author}{\bibfnamefont{E.~T.} \bibnamefont{Neil}},
  \bibinfo{author}{\bibfnamefont{Y.}~\bibnamefont{Shamir}}, \bibnamefont{and}
  \bibinfo{author}{\bibfnamefont{B.}~\bibnamefont{Svetitsky}},
  \bibinfo{journal}{Phys. Rev.} \textbf{\bibinfo{volume}{D97}},
  \bibinfo{pages}{114505} (\bibinfo{year}{2018}{\natexlab{b}}),
  \eprint{1801.05809}.

\bibitem[{\citenamefont{Ayyar et~al.}(2018{\natexlab{c}})\citenamefont{Ayyar,
  DeGrand, Hackett, Jay, Neil, Shamir, and Svetitsky}}]{Ayyar:2018ppa}
\bibinfo{author}{\bibfnamefont{V.}~\bibnamefont{Ayyar}},
  \bibinfo{author}{\bibfnamefont{T.}~\bibnamefont{DeGrand}},
  \bibinfo{author}{\bibfnamefont{D.~C.} \bibnamefont{Hackett}},
  \bibinfo{author}{\bibfnamefont{W.~I.} \bibnamefont{Jay}},
  \bibinfo{author}{\bibfnamefont{E.~T.} \bibnamefont{Neil}},
  \bibinfo{author}{\bibfnamefont{Y.}~\bibnamefont{Shamir}}, \bibnamefont{and}
  \bibinfo{author}{\bibfnamefont{B.}~\bibnamefont{Svetitsky}},
  \bibinfo{journal}{Phys. Rev.} \textbf{\bibinfo{volume}{D97}},
  \bibinfo{pages}{114502} (\bibinfo{year}{2018}{\natexlab{c}}),
  \eprint{1802.09644}.

\bibitem[{\citenamefont{Bennett et~al.}(2018)\citenamefont{Bennett, Hong, Lee,
  Lin, Lucini, Piai, and Vadacchino}}]{Bennett:2017kga}
\bibinfo{author}{\bibfnamefont{E.}~\bibnamefont{Bennett}},
  \bibinfo{author}{\bibfnamefont{D.~K.} \bibnamefont{Hong}},
  \bibinfo{author}{\bibfnamefont{J.-W.} \bibnamefont{Lee}},
  \bibinfo{author}{\bibfnamefont{C.~J.~D.} \bibnamefont{Lin}},
  \bibinfo{author}{\bibfnamefont{B.}~\bibnamefont{Lucini}},
  \bibinfo{author}{\bibfnamefont{M.}~\bibnamefont{Piai}}, \bibnamefont{and}
  \bibinfo{author}{\bibfnamefont{D.}~\bibnamefont{Vadacchino}},
  \bibinfo{journal}{JHEP} \textbf{\bibinfo{volume}{03}}, \bibinfo{pages}{185}
  (\bibinfo{year}{2018}), \eprint{1712.04220}.

\bibitem[{\citenamefont{Lee et~al.}(2018)\citenamefont{Lee, Bennett, Hong, Lin,
  Lucini, Piai, and Vadacchino}}]{Lee:2018ztv}
\bibinfo{author}{\bibfnamefont{J.-W.} \bibnamefont{Lee}},
  \bibinfo{author}{\bibnamefont{Bennett}},
  \bibinfo{author}{\bibfnamefont{D.~K.} \bibnamefont{Hong}},
  \bibinfo{author}{\bibfnamefont{C.~J.~D.} \bibnamefont{Lin}},
  \bibinfo{author}{\bibfnamefont{B.}~\bibnamefont{Lucini}},
  \bibinfo{author}{\bibfnamefont{M.}~\bibnamefont{Piai}}, \bibnamefont{and}
  \bibinfo{author}{\bibfnamefont{D.}~\bibnamefont{Vadacchino}},
  \bibinfo{journal}{PoS} \textbf{\bibinfo{volume}{LATTICE2018}},
  \bibinfo{pages}{192} (\bibinfo{year}{2018}), \eprint{1811.00276}.

\bibitem[{\citenamefont{Hung and Xiong}(2011)}]{Hung:2009hy}
\bibinfo{author}{\bibfnamefont{P.}~\bibnamefont{Hung}} \bibnamefont{and}
  \bibinfo{author}{\bibfnamefont{C.}~\bibnamefont{Xiong}},
  \bibinfo{journal}{Nucl.Phys.} \textbf{\bibinfo{volume}{B847}},
  \bibinfo{pages}{160} (\bibinfo{year}{2011}).

\bibitem[{\citenamefont{Molgaard and Shrock}(2014)}]{Molgaard:2014mqa}
\bibinfo{author}{\bibfnamefont{E.}~\bibnamefont{Molgaard}} \bibnamefont{and}
  \bibinfo{author}{\bibfnamefont{R.}~\bibnamefont{Shrock}},
  \bibinfo{journal}{Phys.Rev.} \textbf{\bibinfo{volume}{D89}},
  \bibinfo{pages}{105007} (\bibinfo{year}{2014}).

\bibitem[{\citenamefont{Shigemitsu}(1987)}]{Shigemitsu:1987ei}
\bibinfo{author}{\bibfnamefont{J.}~\bibnamefont{Shigemitsu}},
  \bibinfo{journal}{Phys.Lett.} \textbf{\bibinfo{volume}{B189}},
  \bibinfo{pages}{164} (\bibinfo{year}{1987}).

\bibitem[{\citenamefont{Lee and Shrock}(1987)}]{Lee:1987eg}
\bibinfo{author}{\bibfnamefont{I.-H.} \bibnamefont{Lee}} \bibnamefont{and}
  \bibinfo{author}{\bibfnamefont{R.~E.} \bibnamefont{Shrock}},
  \bibinfo{journal}{Phys.Rev.Lett.} \textbf{\bibinfo{volume}{59}},
  \bibinfo{pages}{14} (\bibinfo{year}{1987}).

\bibitem[{\citenamefont{Lee and Shrock}(1988)}]{Lee:1988ut}
\bibinfo{author}{\bibfnamefont{I.-H.} \bibnamefont{Lee}} \bibnamefont{and}
  \bibinfo{author}{\bibfnamefont{R.~E.} \bibnamefont{Shrock}},
  \bibinfo{journal}{Nucl.Phys.} \textbf{\bibinfo{volume}{B305}},
  \bibinfo{pages}{305} (\bibinfo{year}{1988}).

\bibitem[{\citenamefont{Bock et~al.}(1989)\citenamefont{Bock, De, Jansen,
  Jersak, and Neuhaus}}]{Bock:1989ye}
\bibinfo{author}{\bibfnamefont{W.}~\bibnamefont{Bock}},
  \bibinfo{author}{\bibfnamefont{A.}~\bibnamefont{De}},
  \bibinfo{author}{\bibfnamefont{K.}~\bibnamefont{Jansen}},
  \bibinfo{author}{\bibfnamefont{J.}~\bibnamefont{Jersak}}, \bibnamefont{and}
  \bibinfo{author}{\bibfnamefont{T.}~\bibnamefont{Neuhaus}},
  \bibinfo{journal}{Phys.Lett.} \textbf{\bibinfo{volume}{B231}},
  \bibinfo{pages}{283} (\bibinfo{year}{1989}).

\bibitem[{\citenamefont{Hasenfratz et~al.}(1990)\citenamefont{Hasenfratz, Liu,
  and Neuhaus}}]{Hasenfratz:1989jr}
\bibinfo{author}{\bibfnamefont{A.}~\bibnamefont{Hasenfratz}},
  \bibinfo{author}{\bibfnamefont{W.-q.} \bibnamefont{Liu}}, \bibnamefont{and}
  \bibinfo{author}{\bibfnamefont{T.}~\bibnamefont{Neuhaus}},
  \bibinfo{journal}{Phys.Lett.} \textbf{\bibinfo{volume}{B236}},
  \bibinfo{pages}{339} (\bibinfo{year}{1990}).

\bibitem[{\citenamefont{Lee et~al.}(1990{\natexlab{a}})\citenamefont{Lee,
  Shigemitsu, and Shrock}}]{Lee:1989xq}
\bibinfo{author}{\bibfnamefont{I.-H.} \bibnamefont{Lee}},
  \bibinfo{author}{\bibfnamefont{J.}~\bibnamefont{Shigemitsu}},
  \bibnamefont{and} \bibinfo{author}{\bibfnamefont{R.~E.}
  \bibnamefont{Shrock}}, \bibinfo{journal}{Nucl.Phys.}
  \textbf{\bibinfo{volume}{B330}}, \bibinfo{pages}{225}
  (\bibinfo{year}{1990}{\natexlab{a}}).

\bibitem[{\citenamefont{Lee et~al.}(1990{\natexlab{b}})\citenamefont{Lee,
  Shigemitsu, and Shrock}}]{Lee:1989mi}
\bibinfo{author}{\bibfnamefont{I.-H.} \bibnamefont{Lee}},
  \bibinfo{author}{\bibfnamefont{J.}~\bibnamefont{Shigemitsu}},
  \bibnamefont{and} \bibinfo{author}{\bibfnamefont{R.~E.}
  \bibnamefont{Shrock}}, \bibinfo{journal}{Nucl.Phys.}
  \textbf{\bibinfo{volume}{B334}}, \bibinfo{pages}{265}
  (\bibinfo{year}{1990}{\natexlab{b}}).

\bibitem[{\citenamefont{Shigemitsu}(1989)}]{Shigemitsu:1989xb}
\bibinfo{author}{\bibfnamefont{J.}~\bibnamefont{Shigemitsu}},
  \bibinfo{journal}{Phys.Lett.} \textbf{\bibinfo{volume}{B226}},
  \bibinfo{pages}{364} (\bibinfo{year}{1989}).

\bibitem[{\citenamefont{Abada and Shrock}(1991)}]{Abada:1990ds}
\bibinfo{author}{\bibfnamefont{A.}~\bibnamefont{Abada}} \bibnamefont{and}
  \bibinfo{author}{\bibfnamefont{R.}~\bibnamefont{Shrock}},
  \bibinfo{journal}{Phys.Rev.} \textbf{\bibinfo{volume}{D43}},
  \bibinfo{pages}{304} (\bibinfo{year}{1991}).

\bibitem[{\citenamefont{Bock et~al.}(1990)}]{Bock:1990tv}
\bibinfo{author}{\bibfnamefont{W.}~\bibnamefont{Bock}} \bibnamefont{et~al.},
  \bibinfo{journal}{Nucl.Phys.} \textbf{\bibinfo{volume}{B344}},
  \bibinfo{pages}{207} (\bibinfo{year}{1990}).

\bibitem[{\citenamefont{Bock et~al.}(1992{\natexlab{a}})\citenamefont{Bock, De,
  Frick, Jansen, and Trappenberg}}]{Bock:1991kn}
\bibinfo{author}{\bibfnamefont{W.}~\bibnamefont{Bock}},
  \bibinfo{author}{\bibfnamefont{A.~K.} \bibnamefont{De}},
  \bibinfo{author}{\bibfnamefont{C.}~\bibnamefont{Frick}},
  \bibinfo{author}{\bibfnamefont{K.}~\bibnamefont{Jansen}}, \bibnamefont{and}
  \bibinfo{author}{\bibfnamefont{T.}~\bibnamefont{Trappenberg}},
  \bibinfo{journal}{Nucl.Phys.} \textbf{\bibinfo{volume}{B371}},
  \bibinfo{pages}{683} (\bibinfo{year}{1992}{\natexlab{a}}).

\bibitem[{\citenamefont{Bock et~al.}(1992{\natexlab{b}})\citenamefont{Bock, De,
  and Smit}}]{Bock:1991bu}
\bibinfo{author}{\bibfnamefont{W.}~\bibnamefont{Bock}},
  \bibinfo{author}{\bibfnamefont{A.~K.} \bibnamefont{De}}, \bibnamefont{and}
  \bibinfo{author}{\bibfnamefont{J.}~\bibnamefont{Smit}},
  \bibinfo{journal}{Nucl.Phys.} \textbf{\bibinfo{volume}{B388}},
  \bibinfo{pages}{243} (\bibinfo{year}{1992}{\natexlab{b}}).

\bibitem[{\citenamefont{Bock et~al.}(1992{\natexlab{c}})\citenamefont{Bock, De,
  Frick, Jersak, and Trappenberg}}]{Bock:1991da}
\bibinfo{author}{\bibfnamefont{W.}~\bibnamefont{Bock}},
  \bibinfo{author}{\bibfnamefont{A.~K.} \bibnamefont{De}},
  \bibinfo{author}{\bibfnamefont{C.}~\bibnamefont{Frick}},
  \bibinfo{author}{\bibfnamefont{J.}~\bibnamefont{Jersak}}, \bibnamefont{and}
  \bibinfo{author}{\bibfnamefont{T.}~\bibnamefont{Trappenberg}},
  \bibinfo{journal}{Nucl.Phys.} \textbf{\bibinfo{volume}{B378}},
  \bibinfo{pages}{652} (\bibinfo{year}{1992}{\natexlab{c}}).

\bibitem[{\citenamefont{Hasenfratz et~al.}(1991)\citenamefont{Hasenfratz,
  Hasenfratz, Jansen, Kuti, and Shen}}]{Hasenfratz:1991it}
\bibinfo{author}{\bibfnamefont{A.}~\bibnamefont{Hasenfratz}},
  \bibinfo{author}{\bibfnamefont{P.}~\bibnamefont{Hasenfratz}},
  \bibinfo{author}{\bibfnamefont{K.}~\bibnamefont{Jansen}},
  \bibinfo{author}{\bibfnamefont{J.}~\bibnamefont{Kuti}}, \bibnamefont{and}
  \bibinfo{author}{\bibfnamefont{Y.}~\bibnamefont{Shen}},
  \bibinfo{journal}{Nucl.Phys.} \textbf{\bibinfo{volume}{B365}},
  \bibinfo{pages}{79} (\bibinfo{year}{1991}).

\bibitem[{\citenamefont{Kuti}()}]{Kuti:1991vg}
\bibinfo{author}{\bibfnamefont{J.}~\bibnamefont{Kuti}}, \bibinfo{note}{{in
  Proceedings of "Electroweak symmetry breaking", Hiroshima 1991}}.

\bibitem[{\citenamefont{Hasenfratz et~al.}(1993)\citenamefont{Hasenfratz,
  Jansen, and Shen}}]{Hasenfratz:1992xs}
\bibinfo{author}{\bibfnamefont{A.}~\bibnamefont{Hasenfratz}},
  \bibinfo{author}{\bibfnamefont{K.}~\bibnamefont{Jansen}}, \bibnamefont{and}
  \bibinfo{author}{\bibfnamefont{Y.}~\bibnamefont{Shen}},
  \bibinfo{journal}{Nucl.Phys.} \textbf{\bibinfo{volume}{B394}},
  \bibinfo{pages}{527} (\bibinfo{year}{1993}).

\bibitem[{\citenamefont{Maiani et~al.}(1992)\citenamefont{Maiani, Testa, and
  Rossi}}]{Maiani:1992pn}
\bibinfo{author}{\bibfnamefont{L.}~\bibnamefont{Maiani}},
  \bibinfo{author}{\bibfnamefont{M.}~\bibnamefont{Testa}}, \bibnamefont{and}
  \bibinfo{author}{\bibfnamefont{G.}~\bibnamefont{Rossi}},
  \bibinfo{journal}{Nucl.Phys.Proc.Suppl.} \textbf{\bibinfo{volume}{29B+C}}
  (\bibinfo{year}{1992}).

\bibitem[{\citenamefont{Fodor et~al.}(2007)\citenamefont{Fodor, Holland, Kuti,
  Nogradi, and Schroeder}}]{Fodor:2007fn}
\bibinfo{author}{\bibfnamefont{Z.}~\bibnamefont{Fodor}},
  \bibinfo{author}{\bibfnamefont{K.}~\bibnamefont{Holland}},
  \bibinfo{author}{\bibfnamefont{J.}~\bibnamefont{Kuti}},
  \bibinfo{author}{\bibfnamefont{D.}~\bibnamefont{Nogradi}}, \bibnamefont{and}
  \bibinfo{author}{\bibfnamefont{C.}~\bibnamefont{Schroeder}},
  \bibinfo{journal}{PoS} \textbf{\bibinfo{volume}{LAT2007}},
  \bibinfo{pages}{056} (\bibinfo{year}{2007}).

\bibitem[{\citenamefont{Gerhold and
  Jansen}(2007{\natexlab{a}})}]{Gerhold:2007yb}
\bibinfo{author}{\bibfnamefont{P.}~\bibnamefont{Gerhold}} \bibnamefont{and}
  \bibinfo{author}{\bibfnamefont{K.}~\bibnamefont{Jansen}},
  \bibinfo{journal}{JHEP} \textbf{\bibinfo{volume}{0709}}, \bibinfo{pages}{041}
  (\bibinfo{year}{2007}{\natexlab{a}}).

\bibitem[{\citenamefont{Gerhold and
  Jansen}(2007{\natexlab{b}})}]{Gerhold:2007gx}
\bibinfo{author}{\bibfnamefont{P.}~\bibnamefont{Gerhold}} \bibnamefont{and}
  \bibinfo{author}{\bibfnamefont{K.}~\bibnamefont{Jansen}},
  \bibinfo{journal}{JHEP} \textbf{\bibinfo{volume}{0710}}, \bibinfo{pages}{001}
  (\bibinfo{year}{2007}{\natexlab{b}}).

\bibitem[{\citenamefont{Gerhold and Jansen}(2009)}]{Gerhold:2009ub}
\bibinfo{author}{\bibfnamefont{P.}~\bibnamefont{Gerhold}} \bibnamefont{and}
  \bibinfo{author}{\bibfnamefont{K.}~\bibnamefont{Jansen}},
  \bibinfo{journal}{JHEP} \textbf{\bibinfo{volume}{0907}}, \bibinfo{pages}{025}
  (\bibinfo{year}{2009}).

\bibitem[{\citenamefont{Gerhold and Jansen}(2010)}]{Gerhold:2010bh}
\bibinfo{author}{\bibfnamefont{P.}~\bibnamefont{Gerhold}} \bibnamefont{and}
  \bibinfo{author}{\bibfnamefont{K.}~\bibnamefont{Jansen}},
  \bibinfo{journal}{JHEP} \textbf{\bibinfo{volume}{1004}}, \bibinfo{pages}{094}
  (\bibinfo{year}{2010}).

\bibitem[{\citenamefont{Gerhold et~al.}(2012)\citenamefont{Gerhold, Jansen, and
  Kallarackal}}]{Gerhold:2011mx}
\bibinfo{author}{\bibfnamefont{P.}~\bibnamefont{Gerhold}},
  \bibinfo{author}{\bibfnamefont{K.}~\bibnamefont{Jansen}}, \bibnamefont{and}
  \bibinfo{author}{\bibfnamefont{J.}~\bibnamefont{Kallarackal}},
  \bibinfo{journal}{Phys.Lett.} \textbf{\bibinfo{volume}{B710}},
  \bibinfo{pages}{697} (\bibinfo{year}{2012}).

\bibitem[{\citenamefont{Bulava et~al.}(2013)}]{Bulava:2012rb}
\bibinfo{author}{\bibfnamefont{J.}~\bibnamefont{Bulava}} \bibnamefont{et~al.},
  \bibinfo{journal}{Adv.High Energy Phys.} \textbf{\bibinfo{volume}{2013}},
  \bibinfo{pages}{875612} (\bibinfo{year}{2013}).

\bibitem[{\citenamefont{Fisher and Barber}(1972)}]{Fisher:1972zza}
\bibinfo{author}{\bibfnamefont{M.~E.} \bibnamefont{Fisher}} \bibnamefont{and}
  \bibinfo{author}{\bibfnamefont{M.~N.} \bibnamefont{Barber}},
  \bibinfo{journal}{Phys.Rev.Lett.} \textbf{\bibinfo{volume}{28}},
  \bibinfo{pages}{1516} (\bibinfo{year}{1972}).

\bibitem[{\citenamefont{Brezin}(1982)}]{Brezin:1981gm}
\bibinfo{author}{\bibfnamefont{E.}~\bibnamefont{Brezin}},
  \bibinfo{journal}{J.Phys.(France)} \textbf{\bibinfo{volume}{43}},
  \bibinfo{pages}{15} (\bibinfo{year}{1982}).

\bibitem[{\citenamefont{Brezin and Zinn-Justin}(1985)}]{Brezin:1985xx}
\bibinfo{author}{\bibfnamefont{E.}~\bibnamefont{Brezin}} \bibnamefont{and}
  \bibinfo{author}{\bibfnamefont{J.}~\bibnamefont{Zinn-Justin}},
  \bibinfo{journal}{Nucl.Phys.} \textbf{\bibinfo{volume}{B257}},
  \bibinfo{pages}{867} (\bibinfo{year}{1985}).

\bibitem[{\citenamefont{Bhattacharjee and Seno}(2001)}]{0305-4470-34-33-302}
\bibinfo{author}{\bibfnamefont{S.~M.} \bibnamefont{Bhattacharjee}}
  \bibnamefont{and} \bibinfo{author}{\bibfnamefont{F.}~\bibnamefont{Seno}},
  \bibinfo{journal}{Journal of Physics A: Mathematical and General}
  \textbf{\bibinfo{volume}{34}}, \bibinfo{pages}{6375} (\bibinfo{year}{2001}),
  \urlprefix\url{http://stacks.iop.org/0305-4470/34/i=33/a=302}.

\bibitem[{\citenamefont{Wolff}(1989)}]{Wolff:1988uh}
\bibinfo{author}{\bibfnamefont{U.}~\bibnamefont{Wolff}},
  \bibinfo{journal}{Phys. Rev. Lett.} \textbf{\bibinfo{volume}{62}},
  \bibinfo{pages}{361} (\bibinfo{year}{1989}).

\bibitem[{\citenamefont{Hasenfratz and Leutwyler}(1990)}]{Hasenfratz:1989pk}
\bibinfo{author}{\bibfnamefont{P.}~\bibnamefont{Hasenfratz}} \bibnamefont{and}
  \bibinfo{author}{\bibfnamefont{H.}~\bibnamefont{Leutwyler}},
  \bibinfo{journal}{Nucl. Phys.} \textbf{\bibinfo{volume}{B343}},
  \bibinfo{pages}{241} (\bibinfo{year}{1990}).

\bibitem[{\citenamefont{Shaposhnikov and
  Wetterich}(2010)}]{Shaposhnikov:2009pv}
\bibinfo{author}{\bibfnamefont{M.}~\bibnamefont{Shaposhnikov}}
  \bibnamefont{and}
  \bibinfo{author}{\bibfnamefont{C.}~\bibnamefont{Wetterich}},
  \bibinfo{journal}{Phys. Lett.} \textbf{\bibinfo{volume}{B683}},
  \bibinfo{pages}{196} (\bibinfo{year}{2010}), \eprint{0912.0208}.

\bibitem[{\citenamefont{Gies et~al.}(2013)\citenamefont{Gies, Rechenberger,
  Scherer, and Zambelli}}]{Gies:2013pma}
\bibinfo{author}{\bibfnamefont{H.}~\bibnamefont{Gies}},
  \bibinfo{author}{\bibfnamefont{S.}~\bibnamefont{Rechenberger}},
  \bibinfo{author}{\bibfnamefont{M.~M.} \bibnamefont{Scherer}},
  \bibnamefont{and} \bibinfo{author}{\bibfnamefont{L.}~\bibnamefont{Zambelli}},
  \bibinfo{journal}{Eur. Phys. J.} \textbf{\bibinfo{volume}{C73}},
  \bibinfo{pages}{2652} (\bibinfo{year}{2013}), \eprint{1306.6508}.

\bibitem[{\citenamefont{Litim and Sannino}(2014)}]{Litim:2014uca}
\bibinfo{author}{\bibfnamefont{D.~F.} \bibnamefont{Litim}} \bibnamefont{and}
  \bibinfo{author}{\bibfnamefont{F.}~\bibnamefont{Sannino}},
  \bibinfo{journal}{JHEP} \textbf{\bibinfo{volume}{12}}, \bibinfo{pages}{178}
  (\bibinfo{year}{2014}), \eprint{1406.2337}.

\bibitem[{\citenamefont{Maas}(2017)}]{Maas:2017wzi}
\bibinfo{author}{\bibfnamefont{A.}~\bibnamefont{Maas}} (\bibinfo{year}{2017}),
  \eprint{1712.04721}.

\bibitem[{\citenamefont{Eichhorn and Held}(2017)}]{Eichhorn:2017eht}
\bibinfo{author}{\bibfnamefont{A.}~\bibnamefont{Eichhorn}} \bibnamefont{and}
  \bibinfo{author}{\bibfnamefont{A.}~\bibnamefont{Held}},
  \bibinfo{journal}{Phys. Rev.} \textbf{\bibinfo{volume}{D96}},
  \bibinfo{pages}{086025} (\bibinfo{year}{2017}), \eprint{1705.02342}.

\end{thebibliography}
 
\end{document}